\documentclass[%
 reprint,
superscriptaddress,
 amsmath,amssymb,
prl,
]{revtex4-2}

\usepackage{graphicx}
\usepackage{dcolumn}
\usepackage{xcolor}
\usepackage{bm}

\newcommand{\siv}{SiV$^{0}$}
\newcommand{\sivm}{SiV$^-$}

\begin{document}

\title{Neutral Silicon Vacancy Centers in Diamond via Photoactivated Itinerant Carriers}
\author{Zi-Huai Zhang}
\affiliation{%
Department of Electrical and Computer Engineering, Princeton University
}%

\author{Andrew M.
Edmonds}
\affiliation{
Element Six, Harwell, OX11 0QR, United Kingdom
}

\author{Nicola Palmer}
\affiliation{
Element Six, Harwell, OX11 0QR, United Kingdom
}

\author{Matthew L. Markham}
\affiliation{
Element Six, Harwell, OX11 0QR, United Kingdom
}

\author{Nathalie P. de Leon}%
 \email{npdeleon@princeton.edu}
\affiliation{%
Department of Electrical and Computer Engineering, Princeton University
}%

\date{\today}

\begin{abstract}
Neutral silicon vacancy (\siv{}) centers in diamond are promising candidates for quantum network applications because of their exceptional optical properties and spin coherence. However, the stabilization of \siv{} centers requires careful Fermi level engineering of the diamond host material, making further technological development challenging. Here, we show that \siv{} centers can be efficiently stabilized by photoactivated itinerant carriers. Even in this nonequilibrium configuration, the resulting \siv{} centers are stable enough to allow for resonant optical excitation and optically detected magnetic resonance. Our results pave the way for on-demand generation of \siv{} centers as well as other emerging quantum defects in diamond.
\end{abstract}

\maketitle
Color centers in the solid state are actively explored as building blocks for quantum sensing and quantum information processing because of their exceptional spin and optical properties \cite{Gao2015,Atature2018,Awschalom2018}. As atomic systems, they can exhibit narrow optical transitions and long spin coherence times. As solid state objects, they can be integrated into devices in a scalable manner. However, the solid state environment also leads to complicated charge dynamics. For example, impurities or dopants in the substrate \cite{Mizuochi2016,Luhmann2019}, optical illumination of photo-active impurities \cite{Aslam2013}, and surface properties of the substrate \cite{Dhomkar2018nl,Yuan2020} can all impact the stability and the dynamics of the desired charge state. Using optical illumination to initialize the charge state is commonly deployed for various color centers, including nitrogen vacancy (NV) and silicon vacancy (SiV) centers in diamond \cite{Beha2012,Siyushev2013,Sipahigil2016}. However, the fidelity of the initialization is limited~\cite{Waldherr2011}, and depends strongly on the microscopic details of the sample, especially when the centers are near the surface~\cite{Dolev2019,Yuan2020}. New methods to control and stabilize particular charge states would enable many technological applications based on color centers.

Recently, it was shown that the charge state of color centers in diamond can be affected by optical illumination of nearby defects \cite{Jayakumar2016}. The optically activated itinerant carriers (holes and electrons) diffuse before they are captured, resulting in a non-local modification of the charge state distribution. Using itinerant carriers that are generated remotely avoids possible complications from ionization and recombination under direct illumination. This procedure has been demonstrated for both NV centers and SiV centers, and the effect has been attributed to optical ionization of nearby NV centers, SiV centers or substitutional nitrogen (P1 centers) \cite{Jayakumar2016,Dhomkar2018,Gardill2021,Lozovoi2021}. For NV centers, the negative charge state (NV$^-$) captures holes efficiently, converting NV$^-$ to NV$^0$ \cite{Gardill2021,Lozovoi2021}. For SiV centers, the negative charge state (\sivm{}) can be generated in a similar manner; however the nature of the charge state prior to conversion is disputed in part because of the lack of direct observation of the neutral silicon vacancy (\siv{}) centers \cite{Dhomkar2018, Gardill2021}.

In this work, we demonstrate stabilization of \siv{} centers in diamond via photoactivation of itinerant carriers. \siv{} centers are unstable in typical high purity diamonds because the Fermi level set by nitrogen impurities is above the charge transition point from the neutral to negative charge states \cite{Gali2013}, and high conversion to the neutral charge state has only been demonstrated in a limited number of boron doped diamonds \cite{Rose2018}, or through surface transfer doping in undoped diamond \cite{Zhang2022}. Here, instead of relying on careful preparation and doping of the diamond, we create a nonequilibrium charge state by utilizing the itinerant carriers generated from nearby defects upon optical illumination. We show that \sivm{} centers are capable of capturing photoactivated holes, which metastably converts them into \siv{} centers. We study the dynamics of carrier diffusion, the ionization of \siv{} under illumination, and the dependence of the dynamics on excitation wavelength. The wavelength dependence allows us to infer that dynamic charge state conversion of NV centers plays the major role in generating the itinerant carriers in our sample. The nonequilibrium \siv{} centers are stable both in the dark and under typical near-infrared (NIR) optical excitation, enabling resonant optical excitation and optically detected magnetic resonance (ODMR). 

The sample studied in this work is a plasma chemical-vapor deposition grown sample doped with isotopically enriched $^{29}$Si during growth (Element Six). We estimate the NV$^-$ concentration to be $\sim$0.03 ppb and the \sivm{} concentration to be $\sim$30 ppb by comparing the photoluminescence signal to the signal of samples with known NV$^-$ and \sivm{} concentrations. The silicon concentration was determined from secondary ion mass spectrometry to be 0.8 ppm. The nitrogen concentration is below the detection limits of Fourier-transform infrared spectroscopy (upper bound of 300~ppb, Fig.~S1~\cite{Supplemental}), but is at least 9 ppb based on the native NV$^-$ concentration~\cite{Edmonds2012}. Experiments were conducted at 10~K in a home built confocal microscope. The confocal microscope has two independent branches with scanning capability for excitation and detection in either the visible range and the NIR range. In the NIR branch, we use 857~nm and 946~nm excitation for \siv{} measurement. In the visible branch, we use 532~nm, 595~nm and 637~nm excitation for carrier generation and \sivm{} measurement \cite{Supplemental}.

\begin{figure}[h!]
  \centering
  \includegraphics[width = 86mm]{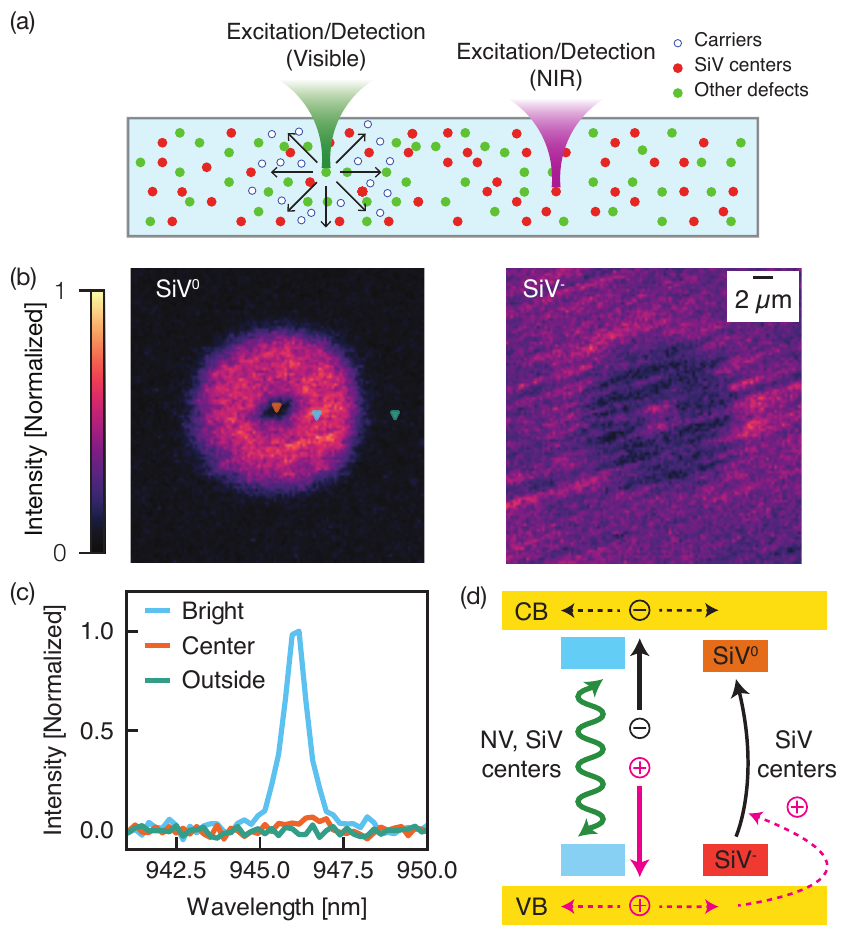}
  \caption{{\bf Photoactivation of \siv{} centers in diamond.}
  (a) Schematic illustration of the measurement setup. A confocal microscope with independent scanning branches for visible and NIR excitation/detection is used for probing SiV centers. Optical illumination using 532 nm produces free carriers and induces a nonequilibrium charge distribution for nearby SiV centers. (b) Confocal fluorescence images detecting \siv{} ($>$930~nm) and \sivm{} (735-745~nm) after 532~nm illumination at the origin (2.58~mW, 60s). A torus shape fluorescence distribution is observed for both \siv{} and \sivm{} centers. (c) Fluorescence spectrum of \siv{} centers at different spatial locations, marked with triangles in (b). Strong \siv{} emission was only observed on the bright torus. (d) Schematic diagram of the proposed charge conversion process. Itinerant carriers are generated via 532~nm illumination of NV and SiV centers in the sample. The carriers diffuse, and holes are captured by \sivm{} centers, converting \sivm{} to \siv{}. }
  \label{fig:Fig1}
\end{figure}

A schematic illustration of the measurement scheme is shown in Fig.~\ref{fig:Fig1}(a). Optical illumination using 532~nm excitation generates itinerant carriers. The photoactivated carriers then diffuse away from the illumination location and get captured by nearby defects. After carrier generation and diffusion, we probe photoluminescence from SiV centers with low power optical excitation. The resulting spatial distribution of SiV centers is shown in Fig.~\ref{fig:Fig1}(b). Four salient features are observable: (1) a bright torus for \siv{} forms around the 532~nm illumination point, (2) the \sivm{} signal shows a dark torus that is inverted compared to \siv{} signal, (3) the torus pattern extends over 10 $\mu$m and SiV charge state remains unaffected outside of the torus, (4) in the center of the torus under direct 532~nm illumination, \siv{} remains dark while \sivm{} remains bright. To confirm that the bright torus originates from \siv{} emission, we measured the optical spectrum at different locations, and emission at the \siv{} zero-phonon line (946~nm) was only observed in the bright torus. By inspecting the SiV charge state far away from 532~nm illumination, we infer that in the sample under study, \sivm{} centers are thermodynamically more favorable compared to \siv{} centers.

The source of itinerant carriers can be impurities and defects such as P1 centers, vacancies, and divacancies, or from color centers such as SiV and NV centers \cite{Jayakumar2016,Dhomkar2018,Gardill2021}. We hypothesize that the observed charge state conversion relies on the charge cycling of color centers (Fig.~\ref{fig:Fig1}(d)), where the simultaneous ionization and recombination processes keep the centers photoactive and generate a continuous flow of itinerant holes and electrons. By contrast, photoionization of a P1 center only produces a single electron and then converts the P1 center to a photo-inactive state. Carriers that are injected into the conduction band and valence band can diffuse away and get captured by defects in the vicinity. We observe that in our sample (Fig.~\ref{fig:Fig1}(b)), the \sivm{} centers capture the free holes to form \siv{} centers while the opposite process (\siv{} centers capturing free electrons) is less likely to happen. This preferential hole capture is consistent with recent observations in NV centers where Coulomb attraction makes the hole capture process for NV$^{-}$ dominant over the electron capture process for NV$^0$ \cite{Jayakumar2016,Lozovoi2021}, and the large hole capture cross section may aris from Rydberg-like bound exciton states in \siv{} \cite{Zhang2020}.

\begin{figure}[h!]
  \centering
  \includegraphics[width = 86mm]{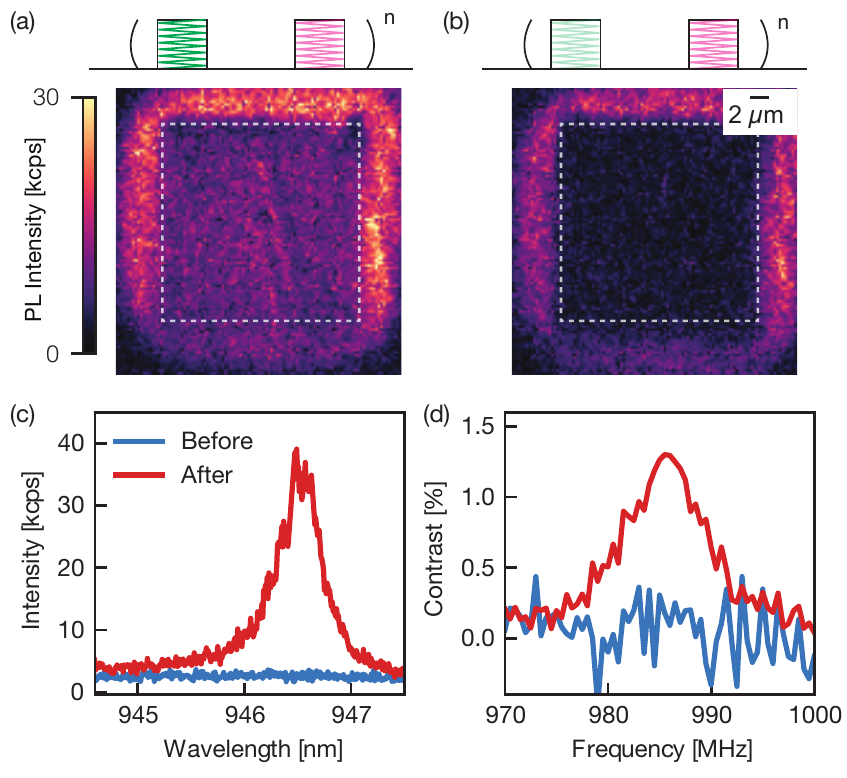}
  \caption{{\bf Initialization of SiV centers and subsequent readout and control of \siv{} centers.}
  (a) A uniform \siv{} rich region is generated by high power (1.1~mW) 532~nm scans. (b) Erasure of \siv{} population by low power (0.23~mW) 532~nm scans. 10 raster scans with 0.2 $\mu$m step size and 1~ms dwell time were used for SiV initialization in (a) and (b). The white dashed boxes denote the excitation scan region. (c) PLE of \siv{} centers before and after photogeneration (300~$\mu$W excitation). The \siv{} population was generated via a finite duration, fixed location 532~nm illumination. (d) Continuous-wave ODMR near the \siv{} zero-field splitting before and after photogeneration. \siv{} was probed 5~$\mu$m away from the 532~nm illumination location. The 532~nm illumination (1.95~mW) was kept on during the measurement. The excitation was set to 855.45~nm with $\sim$11~mW power.}
  \label{fig:Fig2}
\end{figure}

The charge state distribution after a laser raster scan results from the interplay between competing non-local and local effects \cite{Dhomkar2018,Gardill2021}. Utilizing both non-local \siv{} generation and local \sivm{} preparation from 532~nm illumination, we can prepare large areas of SiV centers into desired charge states on-demand. We perform 532~nm initialization scans followed by readout of \siv{} using weak 857~nm excitation. Under high power 532~nm initialization scans, the charge state of SiV is dominated by carrier diffusion and capture, giving rise to higher \siv{} population (Fig.~\ref{fig:Fig2}(a)). Under low power 532~nm initialization scans, the charge state of SiV is dominated by local ionization processes, forming \sivm{} centers (Fig.~\ref{fig:Fig2}(b)). These nonequilibrium \siv{} centers are long-lived in the dark, with no observable decay up to 30 minutes (Fig.~S2 \cite{Supplemental}). We note that \siv{} centers prepared using high power raster scans display lower intensity compared to \siv{} centers generated via only hole capture (Fig.~\ref{fig:Fig2}(a)). This contrast is limited by the non-negligible local ionization dynamics, and demonstrates that higher charge state initialization fidelities can be achieved with itinerant carrier capture. 

The on-demand generation of \siv{} centers stabilizes their charge state in a nonequilibrium configuration. We probe their optical and spin properties to test their stability and utility for quantum applications under illumination. Photoluminescence excitation (PLE) spectroscopy reveals a strong signal at the \siv{} zero-phonon line (946~nm) after optical preparation (Fig.~\ref{fig:Fig2}(c)), as well as an ODMR signal at the zero field splitting (986~MHz) (Fig.~2(d)), while there was no signal before optical illumination. The observed optical and spin transitions of nonequilibrium \siv{} centers are consistent with previous observations \cite{Rose2018,Zhang2020}, demonstrating that charge state stabilization with photoactivated itinerant carriers is an promising alternative approach for generating \siv{}.

\begin{figure}[h!]
  \centering
  \includegraphics[width = 86mm]{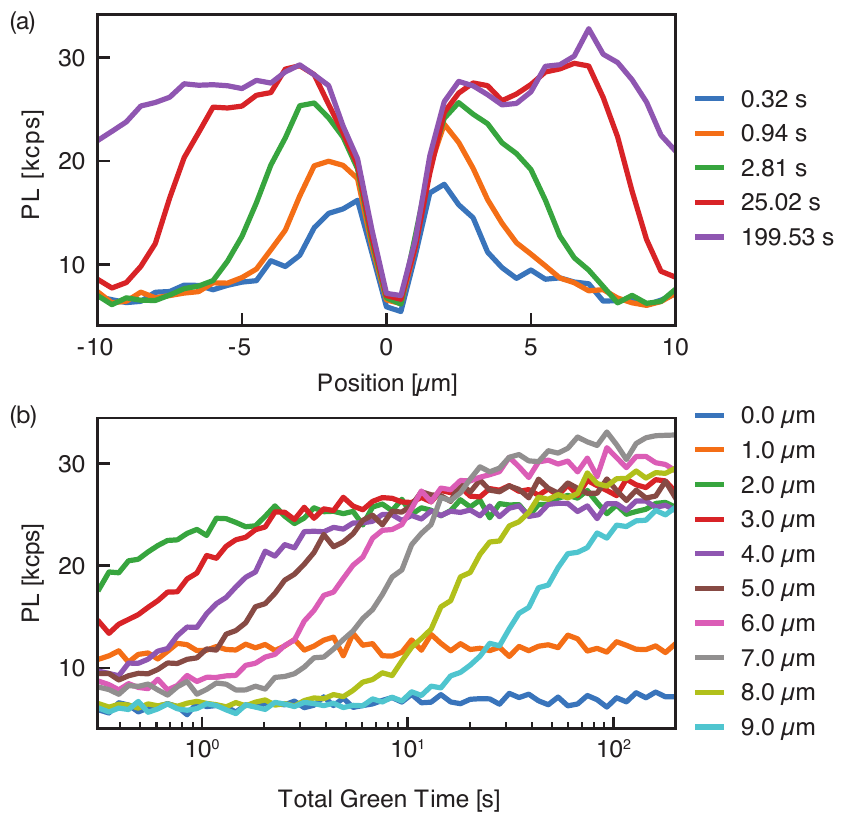}
\caption{{\bf Time dependent \siv{} generation.} (a) Snapshots of the \siv{} distribution after variable 532~nm illumination times. (b) \siv{} population at different positions as a function of illumination time. The local minima of \siv{} population corresponds to the location of 532~nm illumination. The 532~nm laser was set to 1.57~mW. \siv{} was read out using 857~nm (0.29~mW).}
  \label{fig:Fig3}
\end{figure}

We also probe carrier diffusion dynamics and \siv{} formation by taking temporal snapshots of the spatial distribution of \siv{}. The 532~nm illumination and NIR readout are interleaved to measure the time-dependent evolution of \siv{} distribution in a single experimental run. We observe that the size of the torus grows exponentially over time, and the \siv{} intensity saturates at similar levels at long times for different spatial positions (Fig.~\ref{fig:Fig3}), indicating the establishment of a steady-state population. We estimate the charge state initialization to be around 40\% based on the depletion of \sivm{}. We note that the growth rate of the torus appears slower than a diffusion process described by Brownian motion (Fig.~S3 and Fig.~S4~\cite{Supplemental}), which could be related to the high concentration of SiV centers in our sample and space-charge effects \cite{Lozovoi2020}. 

\begin{figure}[h!]
  \centering
  \includegraphics[width = 86mm]{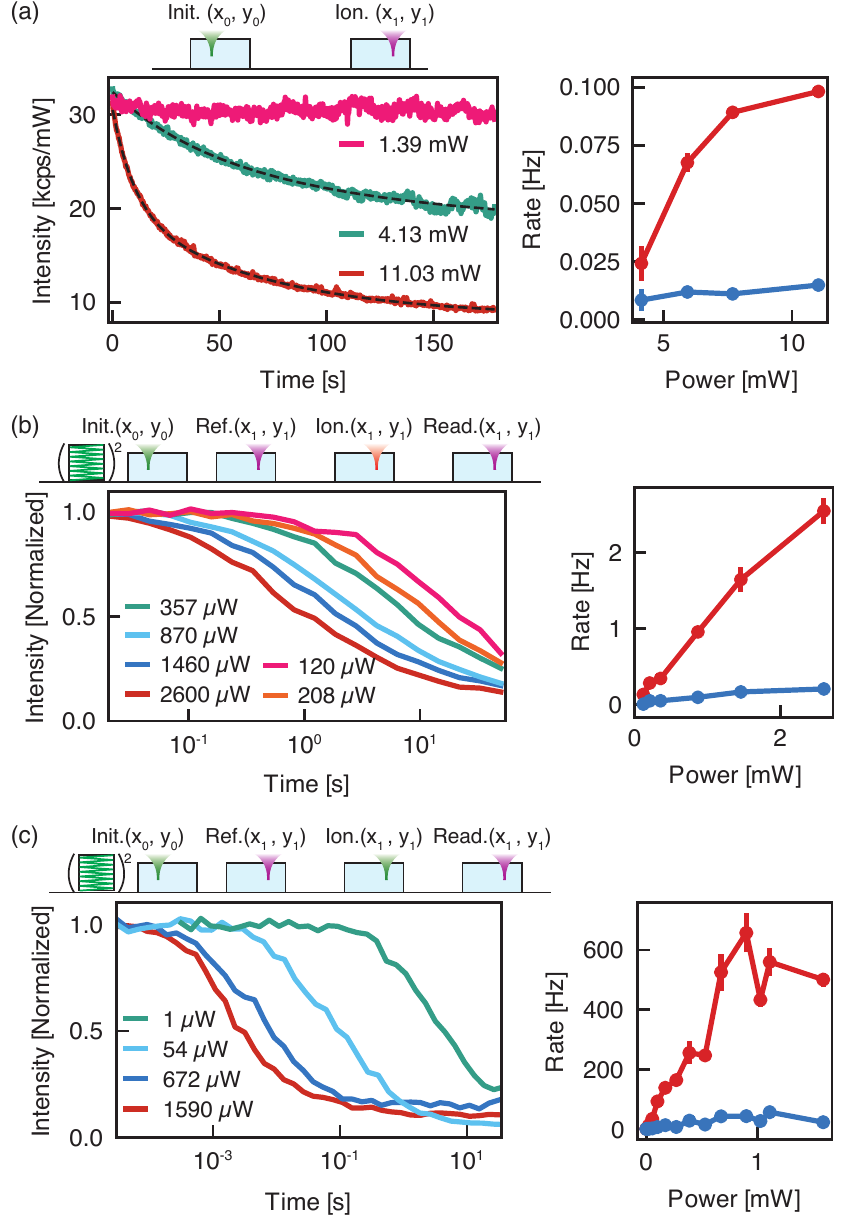}
  \caption{{\bf Ionization dynamics of \siv{} centers under different excitation wavelengths.}
  (a) Ionization of \siv{} under 857~nm excitation. The \siv{} population was monitored in real time. The data is fit to a bi-exponential decay, and the two timescales as a function of power are plotted in the right panel. Ionization of \siv{} under 637 nm excitation (b) and 532 nm excitation (c). For both experiments, \siv{} was prepared via a high power 532~nm scan and high power fixed 532~nm illumination (1-1.5~mW, 5~s), and then ionized with varying duration pulses. A weak 857 nm pulse was used to read out the final \siv{} population, normalized by a reference pulse before ionization. }
  \label{fig:Fig4}
\end{figure}

In order to probe the stability of these nonequilibrium \siv{} centers, we study their ionization dynamics under different excitation wavelengths. For many applications, the relevant excitation wavelength is near the zero phonon line. Using 857~nm excitation, which is below the ionization threshold ($\sim$1.5~eV \cite{Collins1995}), we observe that the \siv{} fluorescence decays over long time scales, with the decay undetectable over minutes for 1.39~mW excitation power (Fig.~\ref{fig:Fig4}(a)). The long-term stability shows that \siv{} centers prepared by transient carrier doping can be stable during optical addressing, over time scales longer than their spin coherence times \cite{Rose2018}. The decay exhibits two characteristic time scales, which has been previously observed in NV and \sivm{} centers \cite{Dhomkar2018nl,Dhomkar2018}. We observe that the faster decay rate increases with power and then saturates at high powers, while the slower rate shows a much weaker power dependence and also appears saturated at higher powers. This power dependence is inconsistent with two-photon ionization of holes to the valence band. One possible microscopic mechanism for the observed saturation is that local charge traps may facilitate ionization, and once the traps are filled, the ionization rate saturates (Fig.~S5(b)~\cite{Supplemental}).

We use 637~nm excitation to study the ionization dynamics above the \siv{} ionization threshold. We observe a decay timescale on the order of seconds (Fig.~\ref{fig:Fig4}(b)). The rate scales linearly with power, consistent with a single-photon ionization process, which is expected for excitation above the ionization threshold.

We also use 532~nm excitation to probe the interplay between the ionization of \siv{} centers and the charge dynamics of the coexisting NV centers. For 532~nm, the ionization rate is much faster, reaching 500 Hz with 1~mW excitation power (Fig.~\ref{fig:Fig4}(c)). The drastically different ionization behavior of \siv{} under 637~nm and 532~nm excitation implies that NV centers play a key role. Specifically, they can act as traps for the transiently generated holes (Fig.~S5(b)~\cite{Supplemental}). Under 532 nm excitation, NV centers are dynamically initialized into NV$^{-}$, and thus can capture holes arising from \siv{} ionization \cite{Louis2019}. Under 637 nm excitation, NV$^-$ centers are ionized into NV$^{0}$. In this case, the NV centers cannot charge cycle and are unable to affect the dynamics of \siv{}. Above 1~mW, the ionization rate saturates. The origin of this saturation is subject to further study, but we note that photoionization of P1 centers under 532~nm can produce a local space-charge potential~\cite{Lozovoi2020}, which may prevent the hole diffusion and limit the \siv{} ionization~\cite{Supplemental}.

\begin{figure}[h!]
  \centering
  \includegraphics[width = 86mm]{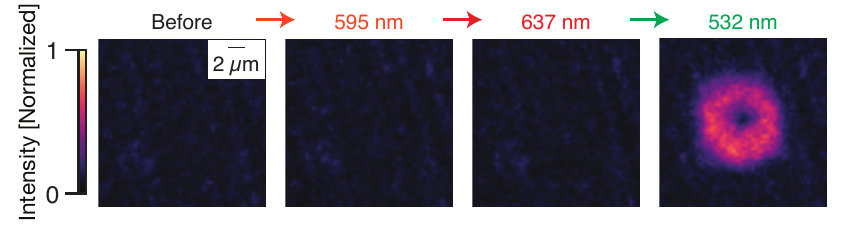}
  \caption{{\bf Wavelength dependence of photogeneration of \siv{}.} Optical illumination at different wavelengths are focused on the same location sequentially for 60~s with 0.34~mW. The arrows indicate the temporal sequence. \siv{} centers were only observed after 532 nm illumination.}
  \label{fig:Fig5}
\end{figure}

Finally, we study the origin of photoactivated carriers in our sample by probing the \siv{} distribution in the same region after laser illumination at different wavelengths. Under the same excitation power and illumination time, we only observed appreciable \siv{} formation with 532~nm but not with 595~nm and 637~nm (Fig.~\ref{fig:Fig5}). For efficient generation of itinerant carriers, the excitation has to enable charge cycling to produce a continuous flow of holes and electrons. For example, NV centers are capable of continuously generating carriers under 532 nm illumination but not under 632 nm illumination because charge cycling is only possible for wavelengths below the zero-phonon line of NV$^0$ at 575~nm \cite{Lozovoi2021}. With the observed sharp transition between 532~nm and 595~nm and the high concentration of NV centers in our sample, we conclude that NV centers play a major role in the observed generation of nonequilibrium \siv{}. We note that we observe a much weaker effect of \siv{} formation under 637~nm excitation at much higher power for an extended duration (Fig.~S7~\cite{Supplemental}), which may be related to optical excitation and charge cycling of SiV centers \cite{Gardill2021}.

We note that recent experiments have reported different charge dynamics for SiV centers. Specifically, \sivm{} centers were observed to enter a dark state under 532~nm illumination or to be converted from a dark state via carrier capture. The nature of this dark state was assigned to either \siv{} or SiV$^{2-}$ \cite{Dhomkar2018,Gardill2021}. In these prior works, the assignment was based on an indirect probe of the dark state where only \sivm{} centers were measured. With the ability to probe both \siv{} and \sivm{} centers, we can assess the charge dynamics more directly. In the sample studied in this manuscript, \sivm{} is dominant in the dark, stable under 532~nm excitation, and can be converted to \siv{} via hole capture. At the same time, in a different sample, we observed signatures of SiV$^{2-}$ in the dark, and sequential formation of \siv{} centers via double hole capture (Fig.~S11~\cite{Supplemental}). We observe that the hole capture photodoping effect of SiV centers is robust across different samples, while the SiV center charge state under direct illumination and the dark state without illumination can be sample dependent, likely due to the different charge environments arising from different impurity concentrations.

In conclusion, we have demonstrated that photoactivated itinerant carriers can be used to stabilize a highly nonequilibrium charge distribution, which opens the possibility of using nonequilibrium charge states of color centers in quantum devices. Specifically, this approach does not rely on complicated optical pulse sequence design or careful bulk substrate engineering. One possible direction is to use photoactivated carriers to stablize single \siv{} centers (Fig.~S12 and Fig.~S13~\cite{Supplemental}) and study charge transport and carrier capture in the presence of an electric field, as has been shown for single NV centers \cite{Lozovoi2021}. Furthermore, our approach for creating nonequilibrium charge distributions should also be applicable to a broader range of diamond color centers. For example, the neutral charge state species of other group IV vacancy centers in diamond remain elusive \cite{Gergo2018,Thiering2019}, and itinerant carrier capture may allow for their stabilization and observation. 

\begin{acknowledgments}
We thank S. Kolkowitz, C. Meriles, and A. Lozovoi for fruitful discussions, as well as L. Rodgers, and Z. Yuan for comments on the manuscript. The study of nonlocal charge state effects was primarily supported by the U.S. Department of Energy, Office of Science, National Quantum Information Science Research Centers, Co-design Center for Quantum Advantage (C2QA) under contract number DE-SC0012704. The confocal microscope and experimental apparatus was built using support from the National Science Foundation through the Princeton Center for Complex Materials, a Materials Research Science and Engineering Center (Grant No. DMR-1420541), and \siv{} charge state spectroscopy was developed with support from the Air Force Office of Scientific Research under Grant No. FA9550-17-0158. 
\end{acknowledgments}

\appendix


\bibliography{doping}

\providecommand{\noopsort}[1]{}\providecommand{\singleletter}[1]{#1}%
\begin{thebibliography}{28}%
\makeatletter
\providecommand \@ifxundefined [1]{%
 \@ifx{#1\undefined}
}%
\providecommand \@ifnum [1]{%
 \ifnum #1\expandafter \@firstoftwo
 \else \expandafter \@secondoftwo
 \fi
}%
\providecommand \@ifx [1]{%
 \ifx #1\expandafter \@firstoftwo
 \else \expandafter \@secondoftwo
 \fi
}%
\providecommand \natexlab [1]{#1}%
\providecommand \enquote  [1]{``#1''}%
\providecommand \bibnamefont  [1]{#1}%
\providecommand \bibfnamefont [1]{#1}%
\providecommand \citenamefont [1]{#1}%
\providecommand \href@noop [0]{\@secondoftwo}%
\providecommand \href [0]{\begingroup \@sanitize@url \@href}%
\providecommand \@href[1]{\@@startlink{#1}\@@href}%
\providecommand \@@href[1]{\endgroup#1\@@endlink}%
\providecommand \@sanitize@url [0]{\catcode `\\12\catcode `\$12\catcode
  `\&12\catcode `\#12\catcode `\^12\catcode `\_12\catcode `\%12\relax}%
\providecommand \@@startlink[1]{}%
\providecommand \@@endlink[0]{}%
\providecommand \url  [0]{\begingroup\@sanitize@url \@url }%
\providecommand \@url [1]{\endgroup\@href {#1}{\urlprefix }}%
\providecommand \urlprefix  [0]{URL }%
\providecommand \Eprint [0]{\href }%
\providecommand \doibase [0]{https://doi.org/}%
\providecommand \selectlanguage [0]{\@gobble}%
\providecommand \bibinfo  [0]{\@secondoftwo}%
\providecommand \bibfield  [0]{\@secondoftwo}%
\providecommand \translation [1]{[#1]}%
\providecommand \BibitemOpen [0]{}%
\providecommand \bibitemStop [0]{}%
\providecommand \bibitemNoStop [0]{.\EOS\space}%
\providecommand \EOS [0]{\spacefactor3000\relax}%
\providecommand \BibitemShut  [1]{\csname bibitem#1\endcsname}%
\let\auto@bib@innerbib\@empty
\bibitem [{\citenamefont {Gao}\ \emph {et~al.}(2015)\citenamefont {Gao},
  \citenamefont {Imamoglu}, \citenamefont {Bernien},\ and\ \citenamefont
  {Hanson}}]{Gao2015}%
  \BibitemOpen
  \bibfield  {author} {\bibinfo {author} {\bibfnamefont {W.~B.}\ \bibnamefont
  {Gao}}, \bibinfo {author} {\bibfnamefont {A.}~\bibnamefont {Imamoglu}},
  \bibinfo {author} {\bibfnamefont {H.}~\bibnamefont {Bernien}},\ and\ \bibinfo
  {author} {\bibfnamefont {R.}~\bibnamefont {Hanson}},\ }\bibfield  {title}
  {\bibinfo {title} {{Coherent manipulation, measurement and entanglement of
  individual solid-state spins using optical fields}},\ }\href
  {https://doi.org/10.1038/nphoton.2015.58} {\bibfield  {journal} {\bibinfo
  {journal} {Nature Photonics}\ }\textbf {\bibinfo {volume} {9}},\ \bibinfo
  {pages} {363} (\bibinfo {year} {2015})}\BibitemShut {NoStop}%
\bibitem [{\citenamefont {Atat{\"{u}}re}\ \emph {et~al.}(2018)\citenamefont
  {Atat{\"{u}}re}, \citenamefont {Englund}, \citenamefont {Vamivakas},
  \citenamefont {Lee},\ and\ \citenamefont {Wrachtrup}}]{Atature2018}%
  \BibitemOpen
  \bibfield  {author} {\bibinfo {author} {\bibfnamefont {M.}~\bibnamefont
  {Atat{\"{u}}re}}, \bibinfo {author} {\bibfnamefont {D.}~\bibnamefont
  {Englund}}, \bibinfo {author} {\bibfnamefont {N.}~\bibnamefont {Vamivakas}},
  \bibinfo {author} {\bibfnamefont {S.-Y.}\ \bibnamefont {Lee}},\ and\ \bibinfo
  {author} {\bibfnamefont {J.}~\bibnamefont {Wrachtrup}},\ }\bibfield  {title}
  {\bibinfo {title} {{Material platforms for spin-based photonic quantum
  technologies}},\ }\href {https://doi.org/10.1038/s41578-018-0008-9}
  {\bibfield  {journal} {\bibinfo  {journal} {Nature Reviews Materials}\
  }\textbf {\bibinfo {volume} {3}},\ \bibinfo {pages} {38} (\bibinfo {year}
  {2018})}\BibitemShut {NoStop}%
\bibitem [{\citenamefont {Awschalom}\ \emph {et~al.}(2018)\citenamefont
  {Awschalom}, \citenamefont {Hanson}, \citenamefont {Wrachtrup},\ and\
  \citenamefont {Zhou}}]{Awschalom2018}%
  \BibitemOpen
  \bibfield  {author} {\bibinfo {author} {\bibfnamefont {D.~D.}\ \bibnamefont
  {Awschalom}}, \bibinfo {author} {\bibfnamefont {R.}~\bibnamefont {Hanson}},
  \bibinfo {author} {\bibfnamefont {J.}~\bibnamefont {Wrachtrup}},\ and\
  \bibinfo {author} {\bibfnamefont {B.~B.}\ \bibnamefont {Zhou}},\ }\bibfield
  {title} {\bibinfo {title} {{Quantum technologies with optically interfaced
  solid-state spins}},\ }\href {https://doi.org/10.1038/s41566-018-0232-2}
  {\bibfield  {journal} {\bibinfo  {journal} {Nature Photonics}\ }\textbf
  {\bibinfo {volume} {12}},\ \bibinfo {pages} {516} (\bibinfo {year}
  {2018})}\BibitemShut {NoStop}%
\bibitem [{\citenamefont {Doi}\ \emph {et~al.}(2016)\citenamefont {Doi},
  \citenamefont {Fukui}, \citenamefont {Kato}, \citenamefont {Makino},
  \citenamefont {Yamasaki}, \citenamefont {Tashima}, \citenamefont {Morishita},
  \citenamefont {Miwa}, \citenamefont {Jelezko}, \citenamefont {Suzuki},\ and\
  \citenamefont {Mizuochi}}]{Mizuochi2016}%
  \BibitemOpen
  \bibfield  {author} {\bibinfo {author} {\bibfnamefont {Y.}~\bibnamefont
  {Doi}}, \bibinfo {author} {\bibfnamefont {T.}~\bibnamefont {Fukui}}, \bibinfo
  {author} {\bibfnamefont {H.}~\bibnamefont {Kato}}, \bibinfo {author}
  {\bibfnamefont {T.}~\bibnamefont {Makino}}, \bibinfo {author} {\bibfnamefont
  {S.}~\bibnamefont {Yamasaki}}, \bibinfo {author} {\bibfnamefont
  {T.}~\bibnamefont {Tashima}}, \bibinfo {author} {\bibfnamefont
  {H.}~\bibnamefont {Morishita}}, \bibinfo {author} {\bibfnamefont
  {S.}~\bibnamefont {Miwa}}, \bibinfo {author} {\bibfnamefont {F.}~\bibnamefont
  {Jelezko}}, \bibinfo {author} {\bibfnamefont {Y.}~\bibnamefont {Suzuki}},\
  and\ \bibinfo {author} {\bibfnamefont {N.}~\bibnamefont {Mizuochi}},\
  }\bibfield  {title} {\bibinfo {title} {Pure negatively charged state of the
  nv center in $n$-type diamond},\ }\href
  {https://doi.org/10.1103/PhysRevB.93.081203} {\bibfield  {journal} {\bibinfo
  {journal} {Phys. Rev. B}\ }\textbf {\bibinfo {volume} {93}},\ \bibinfo
  {pages} {081203} (\bibinfo {year} {2016})}\BibitemShut {NoStop}%
\bibitem [{\citenamefont {L{\"{u}}hmann}\ \emph {et~al.}(2019)\citenamefont
  {L{\"{u}}hmann}, \citenamefont {John}, \citenamefont {Wunderlich},
  \citenamefont {Meijer},\ and\ \citenamefont {Pezzagna}}]{Luhmann2019}%
  \BibitemOpen
  \bibfield  {author} {\bibinfo {author} {\bibfnamefont {T.}~\bibnamefont
  {L{\"{u}}hmann}}, \bibinfo {author} {\bibfnamefont {R.}~\bibnamefont {John}},
  \bibinfo {author} {\bibfnamefont {R.}~\bibnamefont {Wunderlich}}, \bibinfo
  {author} {\bibfnamefont {J.}~\bibnamefont {Meijer}},\ and\ \bibinfo {author}
  {\bibfnamefont {S.}~\bibnamefont {Pezzagna}},\ }\bibfield  {title} {\bibinfo
  {title} {{Coulomb-driven single defect engineering for scalable qubits and
  spin sensors in diamond}},\ }\href
  {https://doi.org/10.1038/s41467-019-12556-0} {\bibfield  {journal} {\bibinfo
  {journal} {Nature Communications}\ }\textbf {\bibinfo {volume} {10}},\
  \bibinfo {pages} {4956} (\bibinfo {year} {2019})}\BibitemShut {NoStop}%
\bibitem [{\citenamefont {Aslam}\ \emph {et~al.}(2013)\citenamefont {Aslam},
  \citenamefont {Waldherr}, \citenamefont {Neumann}, \citenamefont {Jelezko},\
  and\ \citenamefont {Wrachtrup}}]{Aslam2013}%
  \BibitemOpen
  \bibfield  {author} {\bibinfo {author} {\bibfnamefont {N.}~\bibnamefont
  {Aslam}}, \bibinfo {author} {\bibfnamefont {G.}~\bibnamefont {Waldherr}},
  \bibinfo {author} {\bibfnamefont {P.}~\bibnamefont {Neumann}}, \bibinfo
  {author} {\bibfnamefont {F.}~\bibnamefont {Jelezko}},\ and\ \bibinfo {author}
  {\bibfnamefont {J.}~\bibnamefont {Wrachtrup}},\ }\bibfield  {title} {\bibinfo
  {title} {Photo-induced ionization dynamics of the nitrogen vacancy defect in
  diamond investigated by single-shot charge state detection},\ }\href
  {https://doi.org/10.1088/1367-2630/15/1/013064} {\bibfield  {journal}
  {\bibinfo  {journal} {New Journal of Physics}\ }\textbf {\bibinfo {volume}
  {15}},\ \bibinfo {pages} {013064} (\bibinfo {year} {2013})}\BibitemShut
  {NoStop}%
\bibitem [{\citenamefont {Dhomkar}\ \emph
  {et~al.}(2018{\natexlab{a}})\citenamefont {Dhomkar}, \citenamefont
  {Jayakumar}, \citenamefont {Zangara},\ and\ \citenamefont
  {Meriles}}]{Dhomkar2018nl}%
  \BibitemOpen
  \bibfield  {author} {\bibinfo {author} {\bibfnamefont {S.}~\bibnamefont
  {Dhomkar}}, \bibinfo {author} {\bibfnamefont {H.}~\bibnamefont {Jayakumar}},
  \bibinfo {author} {\bibfnamefont {P.~R.}\ \bibnamefont {Zangara}},\ and\
  \bibinfo {author} {\bibfnamefont {C.~A.}\ \bibnamefont {Meriles}},\
  }\bibfield  {title} {\bibinfo {title} {{Charge Dynamics in near-Surface,
  Variable-Density Ensembles of Nitrogen-Vacancy Centers in Diamond}},\ }\href
  {https://doi.org/10.1021/acs.nanolett.8b01739} {\bibfield  {journal}
  {\bibinfo  {journal} {Nano Letters}\ }\textbf {\bibinfo {volume} {18}},\
  \bibinfo {pages} {4046} (\bibinfo {year} {2018}{\natexlab{a}})}\BibitemShut
  {NoStop}%
\bibitem [{\citenamefont {Yuan}\ \emph {et~al.}(2020)\citenamefont {Yuan},
  \citenamefont {Fitzpatrick}, \citenamefont {Rodgers}, \citenamefont
  {Sangtawesin}, \citenamefont {Srinivasan},\ and\ \citenamefont
  {de~Leon}}]{Yuan2020}%
  \BibitemOpen
  \bibfield  {author} {\bibinfo {author} {\bibfnamefont {Z.}~\bibnamefont
  {Yuan}}, \bibinfo {author} {\bibfnamefont {M.}~\bibnamefont {Fitzpatrick}},
  \bibinfo {author} {\bibfnamefont {L.~V.~H.}\ \bibnamefont {Rodgers}},
  \bibinfo {author} {\bibfnamefont {S.}~\bibnamefont {Sangtawesin}}, \bibinfo
  {author} {\bibfnamefont {S.}~\bibnamefont {Srinivasan}},\ and\ \bibinfo
  {author} {\bibfnamefont {N.~P.}\ \bibnamefont {de~Leon}},\ }\bibfield
  {title} {\bibinfo {title} {Charge state dynamics and optically detected
  electron spin resonance contrast of shallow nitrogen-vacancy centers in
  diamond},\ }\href {https://doi.org/10.1103/PhysRevResearch.2.033263}
  {\bibfield  {journal} {\bibinfo  {journal} {Phys. Rev. Research}\ }\textbf
  {\bibinfo {volume} {2}},\ \bibinfo {pages} {033263} (\bibinfo {year}
  {2020})}\BibitemShut {NoStop}%
\bibitem [{\citenamefont {Beha}\ \emph {et~al.}(2012)\citenamefont {Beha},
  \citenamefont {Batalov}, \citenamefont {Manson}, \citenamefont
  {Bratschitsch},\ and\ \citenamefont {Leitenstorfer}}]{Beha2012}%
  \BibitemOpen
  \bibfield  {author} {\bibinfo {author} {\bibfnamefont {K.}~\bibnamefont
  {Beha}}, \bibinfo {author} {\bibfnamefont {A.}~\bibnamefont {Batalov}},
  \bibinfo {author} {\bibfnamefont {N.~B.}\ \bibnamefont {Manson}}, \bibinfo
  {author} {\bibfnamefont {R.}~\bibnamefont {Bratschitsch}},\ and\ \bibinfo
  {author} {\bibfnamefont {A.}~\bibnamefont {Leitenstorfer}},\ }\bibfield
  {title} {\bibinfo {title} {Optimum photoluminescence excitation and
  recharging cycle of single nitrogen-vacancy centers in ultrapure diamond},\
  }\href {https://doi.org/10.1103/PhysRevLett.109.097404} {\bibfield  {journal}
  {\bibinfo  {journal} {Phys. Rev. Lett.}\ }\textbf {\bibinfo {volume} {109}},\
  \bibinfo {pages} {097404} (\bibinfo {year} {2012})}\BibitemShut {NoStop}%
\bibitem [{\citenamefont {Siyushev}\ \emph {et~al.}(2013)\citenamefont
  {Siyushev}, \citenamefont {Pinto}, \citenamefont {V\"or\"os}, \citenamefont
  {Gali}, \citenamefont {Jelezko},\ and\ \citenamefont
  {Wrachtrup}}]{Siyushev2013}%
  \BibitemOpen
  \bibfield  {author} {\bibinfo {author} {\bibfnamefont {P.}~\bibnamefont
  {Siyushev}}, \bibinfo {author} {\bibfnamefont {H.}~\bibnamefont {Pinto}},
  \bibinfo {author} {\bibfnamefont {M.}~\bibnamefont {V\"or\"os}}, \bibinfo
  {author} {\bibfnamefont {A.}~\bibnamefont {Gali}}, \bibinfo {author}
  {\bibfnamefont {F.}~\bibnamefont {Jelezko}},\ and\ \bibinfo {author}
  {\bibfnamefont {J.}~\bibnamefont {Wrachtrup}},\ }\bibfield  {title} {\bibinfo
  {title} {Optically controlled switching of the charge state of a single
  nitrogen-vacancy center in diamond at cryogenic temperatures},\ }\href
  {https://doi.org/10.1103/PhysRevLett.110.167402} {\bibfield  {journal}
  {\bibinfo  {journal} {Phys. Rev. Lett.}\ }\textbf {\bibinfo {volume} {110}},\
  \bibinfo {pages} {167402} (\bibinfo {year} {2013})}\BibitemShut {NoStop}%
\bibitem [{\citenamefont {Sipahigil}\ \emph {et~al.}(2016)\citenamefont
  {Sipahigil}, \citenamefont {Evans}, \citenamefont {Sukachev}, \citenamefont
  {Burek}, \citenamefont {Borregaard}, \citenamefont {Bhaskar}, \citenamefont
  {Nguyen}, \citenamefont {Pacheco}, \citenamefont {Atikian}, \citenamefont
  {Meuwly}, \citenamefont {Camacho}, \citenamefont {Jelezko}, \citenamefont
  {Bielejec}, \citenamefont {Park}, \citenamefont {Lončar},\ and\
  \citenamefont {Lukin}}]{Sipahigil2016}%
  \BibitemOpen
  \bibfield  {author} {\bibinfo {author} {\bibfnamefont {A.}~\bibnamefont
  {Sipahigil}}, \bibinfo {author} {\bibfnamefont {R.~E.}\ \bibnamefont
  {Evans}}, \bibinfo {author} {\bibfnamefont {D.~D.}\ \bibnamefont {Sukachev}},
  \bibinfo {author} {\bibfnamefont {M.~J.}\ \bibnamefont {Burek}}, \bibinfo
  {author} {\bibfnamefont {J.}~\bibnamefont {Borregaard}}, \bibinfo {author}
  {\bibfnamefont {M.~K.}\ \bibnamefont {Bhaskar}}, \bibinfo {author}
  {\bibfnamefont {C.~T.}\ \bibnamefont {Nguyen}}, \bibinfo {author}
  {\bibfnamefont {J.~L.}\ \bibnamefont {Pacheco}}, \bibinfo {author}
  {\bibfnamefont {H.~A.}\ \bibnamefont {Atikian}}, \bibinfo {author}
  {\bibfnamefont {C.}~\bibnamefont {Meuwly}}, \bibinfo {author} {\bibfnamefont
  {R.~M.}\ \bibnamefont {Camacho}}, \bibinfo {author} {\bibfnamefont
  {F.}~\bibnamefont {Jelezko}}, \bibinfo {author} {\bibfnamefont
  {E.}~\bibnamefont {Bielejec}}, \bibinfo {author} {\bibfnamefont
  {H.}~\bibnamefont {Park}}, \bibinfo {author} {\bibfnamefont {M.}~\bibnamefont
  {Lončar}},\ and\ \bibinfo {author} {\bibfnamefont {M.~D.}\ \bibnamefont
  {Lukin}},\ }\bibfield  {title} {\bibinfo {title} {An integrated diamond
  nanophotonics platform for quantum-optical networks},\ }\href
  {https://doi.org/10.1126/science.aah6875} {\bibfield  {journal} {\bibinfo
  {journal} {Science}\ }\textbf {\bibinfo {volume} {354}},\ \bibinfo {pages}
  {847} (\bibinfo {year} {2016})}\BibitemShut {NoStop}%
\bibitem [{\citenamefont {Waldherr}\ \emph {et~al.}(2011)\citenamefont
  {Waldherr}, \citenamefont {Beck}, \citenamefont {Steiner}, \citenamefont
  {Neumann}, \citenamefont {Gali}, \citenamefont {Frauenheim}, \citenamefont
  {Jelezko},\ and\ \citenamefont {Wrachtrup}}]{Waldherr2011}%
  \BibitemOpen
  \bibfield  {author} {\bibinfo {author} {\bibfnamefont {G.}~\bibnamefont
  {Waldherr}}, \bibinfo {author} {\bibfnamefont {J.}~\bibnamefont {Beck}},
  \bibinfo {author} {\bibfnamefont {M.}~\bibnamefont {Steiner}}, \bibinfo
  {author} {\bibfnamefont {P.}~\bibnamefont {Neumann}}, \bibinfo {author}
  {\bibfnamefont {A.}~\bibnamefont {Gali}}, \bibinfo {author} {\bibfnamefont
  {T.}~\bibnamefont {Frauenheim}}, \bibinfo {author} {\bibfnamefont
  {F.}~\bibnamefont {Jelezko}},\ and\ \bibinfo {author} {\bibfnamefont
  {J.}~\bibnamefont {Wrachtrup}},\ }\bibfield  {title} {\bibinfo {title} {{Dark
  States of Single Nitrogen-Vacancy Centers in Diamond Unraveled by Single Shot
  NMR}},\ }\href {https://doi.org/10.1103/PhysRevLett.106.157601} {\bibfield
  {journal} {\bibinfo  {journal} {Phys. Rev. Lett.}\ }\textbf {\bibinfo
  {volume} {106}},\ \bibinfo {pages} {157601} (\bibinfo {year}
  {2011})}\BibitemShut {NoStop}%
\bibitem [{\citenamefont {Bluvstein}\ \emph {et~al.}(2019)\citenamefont
  {Bluvstein}, \citenamefont {Zhang},\ and\ \citenamefont
  {Jayich}}]{Dolev2019}%
  \BibitemOpen
  \bibfield  {author} {\bibinfo {author} {\bibfnamefont {D.}~\bibnamefont
  {Bluvstein}}, \bibinfo {author} {\bibfnamefont {Z.}~\bibnamefont {Zhang}},\
  and\ \bibinfo {author} {\bibfnamefont {A.~C.~B.}\ \bibnamefont {Jayich}},\
  }\bibfield  {title} {\bibinfo {title} {Identifying and mitigating charge
  instabilities in shallow diamond nitrogen-vacancy centers},\ }\href
  {https://doi.org/10.1103/PhysRevLett.122.076101} {\bibfield  {journal}
  {\bibinfo  {journal} {Phys. Rev. Lett.}\ }\textbf {\bibinfo {volume} {122}},\
  \bibinfo {pages} {076101} (\bibinfo {year} {2019})}\BibitemShut {NoStop}%
\bibitem [{\citenamefont {Jayakumar}\ \emph {et~al.}(2016)\citenamefont
  {Jayakumar}, \citenamefont {Henshaw}, \citenamefont {Dhomkar}, \citenamefont
  {Pagliero}, \citenamefont {Laraoui}, \citenamefont {Manson}, \citenamefont
  {Albu}, \citenamefont {Doherty},\ and\ \citenamefont
  {Meriles}}]{Jayakumar2016}%
  \BibitemOpen
  \bibfield  {author} {\bibinfo {author} {\bibfnamefont {H.}~\bibnamefont
  {Jayakumar}}, \bibinfo {author} {\bibfnamefont {J.}~\bibnamefont {Henshaw}},
  \bibinfo {author} {\bibfnamefont {S.}~\bibnamefont {Dhomkar}}, \bibinfo
  {author} {\bibfnamefont {D.}~\bibnamefont {Pagliero}}, \bibinfo {author}
  {\bibfnamefont {A.}~\bibnamefont {Laraoui}}, \bibinfo {author} {\bibfnamefont
  {N.~B.}\ \bibnamefont {Manson}}, \bibinfo {author} {\bibfnamefont
  {R.}~\bibnamefont {Albu}}, \bibinfo {author} {\bibfnamefont {M.~W.}\
  \bibnamefont {Doherty}},\ and\ \bibinfo {author} {\bibfnamefont {C.~A.}\
  \bibnamefont {Meriles}},\ }\bibfield  {title} {\bibinfo {title} {{Optical
  patterning of trapped charge in nitrogen-doped diamond}},\ }\href
  {https://doi.org/10.1038/ncomms12660} {\bibfield  {journal} {\bibinfo
  {journal} {Nature Communications}\ }\textbf {\bibinfo {volume} {7}},\
  \bibinfo {pages} {12660} (\bibinfo {year} {2016})}\BibitemShut {NoStop}%
\bibitem [{\citenamefont {Dhomkar}\ \emph
  {et~al.}(2018{\natexlab{b}})\citenamefont {Dhomkar}, \citenamefont {Zangara},
  \citenamefont {Henshaw},\ and\ \citenamefont {Meriles}}]{Dhomkar2018}%
  \BibitemOpen
  \bibfield  {author} {\bibinfo {author} {\bibfnamefont {S.}~\bibnamefont
  {Dhomkar}}, \bibinfo {author} {\bibfnamefont {P.~R.}\ \bibnamefont
  {Zangara}}, \bibinfo {author} {\bibfnamefont {J.}~\bibnamefont {Henshaw}},\
  and\ \bibinfo {author} {\bibfnamefont {C.~A.}\ \bibnamefont {Meriles}},\
  }\bibfield  {title} {\bibinfo {title} {On-demand generation of neutral and
  negatively charged silicon-vacancy centers in diamond},\ }\href
  {https://doi.org/10.1103/PhysRevLett.120.117401} {\bibfield  {journal}
  {\bibinfo  {journal} {Phys. Rev. Lett.}\ }\textbf {\bibinfo {volume} {120}},\
  \bibinfo {pages} {117401} (\bibinfo {year} {2018}{\natexlab{b}})}\BibitemShut
  {NoStop}%
\bibitem [{\citenamefont {Gardill}\ \emph {et~al.}(2021)\citenamefont
  {Gardill}, \citenamefont {Kemeny}, \citenamefont {Cambria}, \citenamefont
  {Li}, \citenamefont {Dinani}, \citenamefont {Norambuena}, \citenamefont
  {Maze}, \citenamefont {Lordi},\ and\ \citenamefont
  {Kolkowitz}}]{Gardill2021}%
  \BibitemOpen
  \bibfield  {author} {\bibinfo {author} {\bibfnamefont {A.}~\bibnamefont
  {Gardill}}, \bibinfo {author} {\bibfnamefont {I.}~\bibnamefont {Kemeny}},
  \bibinfo {author} {\bibfnamefont {M.~C.}\ \bibnamefont {Cambria}}, \bibinfo
  {author} {\bibfnamefont {Y.}~\bibnamefont {Li}}, \bibinfo {author}
  {\bibfnamefont {H.~T.}\ \bibnamefont {Dinani}}, \bibinfo {author}
  {\bibfnamefont {A.}~\bibnamefont {Norambuena}}, \bibinfo {author}
  {\bibfnamefont {J.~R.}\ \bibnamefont {Maze}}, \bibinfo {author}
  {\bibfnamefont {V.}~\bibnamefont {Lordi}},\ and\ \bibinfo {author}
  {\bibfnamefont {S.}~\bibnamefont {Kolkowitz}},\ }\bibfield  {title} {\bibinfo
  {title} {{Probing Charge Dynamics in Diamond with an Individual Color
  Center}},\ }\href {https://doi.org/10.1021/acs.nanolett.1c02250} {\bibfield
  {journal} {\bibinfo  {journal} {Nano Letters}\ }\textbf {\bibinfo {volume}
  {21}},\ \bibinfo {pages} {6960} (\bibinfo {year} {2021})}\BibitemShut
  {NoStop}%
\bibitem [{\citenamefont {Lozovoi}\ \emph {et~al.}(2021)\citenamefont
  {Lozovoi}, \citenamefont {Jayakumar}, \citenamefont {Daw}, \citenamefont
  {Vizkelethy}, \citenamefont {Bielejec}, \citenamefont {Doherty},
  \citenamefont {Flick},\ and\ \citenamefont {Meriles}}]{Lozovoi2021}%
  \BibitemOpen
  \bibfield  {author} {\bibinfo {author} {\bibfnamefont {A.}~\bibnamefont
  {Lozovoi}}, \bibinfo {author} {\bibfnamefont {H.}~\bibnamefont {Jayakumar}},
  \bibinfo {author} {\bibfnamefont {D.}~\bibnamefont {Daw}}, \bibinfo {author}
  {\bibfnamefont {G.}~\bibnamefont {Vizkelethy}}, \bibinfo {author}
  {\bibfnamefont {E.}~\bibnamefont {Bielejec}}, \bibinfo {author}
  {\bibfnamefont {M.~W.}\ \bibnamefont {Doherty}}, \bibinfo {author}
  {\bibfnamefont {J.}~\bibnamefont {Flick}},\ and\ \bibinfo {author}
  {\bibfnamefont {C.~A.}\ \bibnamefont {Meriles}},\ }\bibfield  {title}
  {\bibinfo {title} {{Optical activation and detection of charge transport
  between individual colour centres in diamond}},\ }\href
  {https://doi.org/10.1038/s41928-021-00656-z} {\bibfield  {journal} {\bibinfo
  {journal} {Nature Electronics}\ }\textbf {\bibinfo {volume} {4}},\ \bibinfo
  {pages} {717} (\bibinfo {year} {2021})}\BibitemShut {NoStop}%
\bibitem [{\citenamefont {Gali}\ and\ \citenamefont {Maze}(2013)}]{Gali2013}%
  \BibitemOpen
  \bibfield  {author} {\bibinfo {author} {\bibfnamefont {A.}~\bibnamefont
  {Gali}}\ and\ \bibinfo {author} {\bibfnamefont {J.~R.}\ \bibnamefont
  {Maze}},\ }\bibfield  {title} {\bibinfo {title} {Ab initio study of the split
  silicon-vacancy defect in diamond: Electronic structure and related
  properties},\ }\href {https://doi.org/10.1103/PhysRevB.88.235205} {\bibfield
  {journal} {\bibinfo  {journal} {Phys. Rev. B}\ }\textbf {\bibinfo {volume}
  {88}},\ \bibinfo {pages} {235205} (\bibinfo {year} {2013})}\BibitemShut
  {NoStop}%
\bibitem [{\citenamefont {Rose}\ \emph {et~al.}(2018)\citenamefont {Rose},
  \citenamefont {Huang}, \citenamefont {Zhang}, \citenamefont {Stevenson},
  \citenamefont {Tyryshkin}, \citenamefont {Sangtawesin}, \citenamefont
  {Srinivasan}, \citenamefont {Loudin}, \citenamefont {Markham}, \citenamefont
  {Edmonds}, \citenamefont {Twitchen}, \citenamefont {Lyon},\ and\
  \citenamefont {de~Leon}}]{Rose2018}%
  \BibitemOpen
  \bibfield  {author} {\bibinfo {author} {\bibfnamefont {B.~C.}\ \bibnamefont
  {Rose}}, \bibinfo {author} {\bibfnamefont {D.}~\bibnamefont {Huang}},
  \bibinfo {author} {\bibfnamefont {Z.-H.}\ \bibnamefont {Zhang}}, \bibinfo
  {author} {\bibfnamefont {P.}~\bibnamefont {Stevenson}}, \bibinfo {author}
  {\bibfnamefont {A.~M.}\ \bibnamefont {Tyryshkin}}, \bibinfo {author}
  {\bibfnamefont {S.}~\bibnamefont {Sangtawesin}}, \bibinfo {author}
  {\bibfnamefont {S.}~\bibnamefont {Srinivasan}}, \bibinfo {author}
  {\bibfnamefont {L.}~\bibnamefont {Loudin}}, \bibinfo {author} {\bibfnamefont
  {M.~L.}\ \bibnamefont {Markham}}, \bibinfo {author} {\bibfnamefont {A.~M.}\
  \bibnamefont {Edmonds}}, \bibinfo {author} {\bibfnamefont {D.~J.}\
  \bibnamefont {Twitchen}}, \bibinfo {author} {\bibfnamefont {S.~A.}\
  \bibnamefont {Lyon}},\ and\ \bibinfo {author} {\bibfnamefont {N.~P.}\
  \bibnamefont {de~Leon}},\ }\bibfield  {title} {\bibinfo {title} {Observation
  of an environmentally insensitive solid-state spin defect in diamond},\
  }\href {https://doi.org/10.1126/science.aao0290} {\bibfield  {journal}
  {\bibinfo  {journal} {Science}\ }\textbf {\bibinfo {volume} {361}},\ \bibinfo
  {pages} {60} (\bibinfo {year} {2018})}\BibitemShut {NoStop}%
\bibitem [{\citenamefont {Zhang}\ \emph {et~al.}(2022)\citenamefont {Zhang},
  \citenamefont {Zuber}, \citenamefont {Rodgers}, \citenamefont {Gui},
  \citenamefont {Stevenson}, \citenamefont {Li}, \citenamefont {Batzer},
  \citenamefont {Grimau}, \citenamefont {Shields}, \citenamefont {Edmonds},
  \citenamefont {Palmer}, \citenamefont {Markham}, \citenamefont {Cava},
  \citenamefont {Maletinsky},\ and\ \citenamefont {de~Leon}}]{Zhang2022}%
  \BibitemOpen
  \bibfield  {author} {\bibinfo {author} {\bibfnamefont {Z.-H.}\ \bibnamefont
  {Zhang}}, \bibinfo {author} {\bibfnamefont {J.~A.}\ \bibnamefont {Zuber}},
  \bibinfo {author} {\bibfnamefont {L.~V.~H.}\ \bibnamefont {Rodgers}},
  \bibinfo {author} {\bibfnamefont {X.}~\bibnamefont {Gui}}, \bibinfo {author}
  {\bibfnamefont {P.}~\bibnamefont {Stevenson}}, \bibinfo {author}
  {\bibfnamefont {M.}~\bibnamefont {Li}}, \bibinfo {author} {\bibfnamefont
  {M.}~\bibnamefont {Batzer}}, \bibinfo {author} {\bibfnamefont {M.~l.}\
  \bibnamefont {Grimau}}, \bibinfo {author} {\bibfnamefont {B.}~\bibnamefont
  {Shields}}, \bibinfo {author} {\bibfnamefont {A.~M.}\ \bibnamefont
  {Edmonds}}, \bibinfo {author} {\bibfnamefont {N.}~\bibnamefont {Palmer}},
  \bibinfo {author} {\bibfnamefont {M.~L.}\ \bibnamefont {Markham}}, \bibinfo
  {author} {\bibfnamefont {R.~J.}\ \bibnamefont {Cava}}, \bibinfo {author}
  {\bibfnamefont {P.}~\bibnamefont {Maletinsky}},\ and\ \bibinfo {author}
  {\bibfnamefont {N.~P.}\ \bibnamefont {de~Leon}},\ }\bibfield  {title}
  {\bibinfo {title} {Neutral silicon vacancy centers in undoped diamond via
  surface control},\ }\bibfield  {journal} {\bibinfo  {journal} {arXiv}\ }\href
  {https://doi.org/10.48550/ARXIV.2206.13698} {10.48550/ARXIV.2206.13698}
  (\bibinfo {year} {2022})\BibitemShut {NoStop}%
\bibitem [{Sup()}]{Supplemental}%
  \BibitemOpen
  \href@noop {} {}\bibinfo {note} {See Supplemental Material for methods and
  additional characterization data.}\BibitemShut {Stop}%
\bibitem [{\citenamefont {Edmonds}\ \emph {et~al.}(2012)\citenamefont
  {Edmonds}, \citenamefont {D'Haenens-Johansson}, \citenamefont {Cruddace},
  \citenamefont {Newton}, \citenamefont {Fu}, \citenamefont {Santori},
  \citenamefont {Beausoleil}, \citenamefont {Twitchen},\ and\ \citenamefont
  {Markham}}]{Edmonds2012}%
  \BibitemOpen
  \bibfield  {author} {\bibinfo {author} {\bibfnamefont {A.~M.}\ \bibnamefont
  {Edmonds}}, \bibinfo {author} {\bibfnamefont {U.~F.~S.}\ \bibnamefont
  {D'Haenens-Johansson}}, \bibinfo {author} {\bibfnamefont {R.~J.}\
  \bibnamefont {Cruddace}}, \bibinfo {author} {\bibfnamefont {M.~E.}\
  \bibnamefont {Newton}}, \bibinfo {author} {\bibfnamefont {K.-M.~C.}\
  \bibnamefont {Fu}}, \bibinfo {author} {\bibfnamefont {C.}~\bibnamefont
  {Santori}}, \bibinfo {author} {\bibfnamefont {R.~G.}\ \bibnamefont
  {Beausoleil}}, \bibinfo {author} {\bibfnamefont {D.~J.}\ \bibnamefont
  {Twitchen}},\ and\ \bibinfo {author} {\bibfnamefont {M.~L.}\ \bibnamefont
  {Markham}},\ }\bibfield  {title} {\bibinfo {title} {Production of oriented
  nitrogen-vacancy color centers in synthetic diamond},\ }\href
  {https://doi.org/10.1103/PhysRevB.86.035201} {\bibfield  {journal} {\bibinfo
  {journal} {Phys. Rev. B}\ }\textbf {\bibinfo {volume} {86}},\ \bibinfo
  {pages} {035201} (\bibinfo {year} {2012})}\BibitemShut {NoStop}%
\bibitem [{\citenamefont {Zhang}\ \emph {et~al.}(2020)\citenamefont {Zhang},
  \citenamefont {Stevenson}, \citenamefont {Thiering}, \citenamefont {Rose},
  \citenamefont {Huang}, \citenamefont {Edmonds}, \citenamefont {Markham},
  \citenamefont {Lyon}, \citenamefont {Gali},\ and\ \citenamefont
  {de~Leon}}]{Zhang2020}%
  \BibitemOpen
  \bibfield  {author} {\bibinfo {author} {\bibfnamefont {Z.-H.}\ \bibnamefont
  {Zhang}}, \bibinfo {author} {\bibfnamefont {P.}~\bibnamefont {Stevenson}},
  \bibinfo {author} {\bibfnamefont {G.~m.~H.}\ \bibnamefont {Thiering}},
  \bibinfo {author} {\bibfnamefont {B.~C.}\ \bibnamefont {Rose}}, \bibinfo
  {author} {\bibfnamefont {D.}~\bibnamefont {Huang}}, \bibinfo {author}
  {\bibfnamefont {A.~M.}\ \bibnamefont {Edmonds}}, \bibinfo {author}
  {\bibfnamefont {M.~L.}\ \bibnamefont {Markham}}, \bibinfo {author}
  {\bibfnamefont {S.~A.}\ \bibnamefont {Lyon}}, \bibinfo {author}
  {\bibfnamefont {A.}~\bibnamefont {Gali}},\ and\ \bibinfo {author}
  {\bibfnamefont {N.~P.}\ \bibnamefont {de~Leon}},\ }\bibfield  {title}
  {\bibinfo {title} {Optically detected magnetic resonance in neutral silicon
  vacancy centers in diamond via bound exciton states},\ }\href
  {https://doi.org/10.1103/PhysRevLett.125.237402} {\bibfield  {journal}
  {\bibinfo  {journal} {Phys. Rev. Lett.}\ }\textbf {\bibinfo {volume} {125}},\
  \bibinfo {pages} {237402} (\bibinfo {year} {2020})}\BibitemShut {NoStop}%
\bibitem [{\citenamefont {Lozovoi}\ \emph {et~al.}(2020)\citenamefont
  {Lozovoi}, \citenamefont {Jayakumar}, \citenamefont {Daw}, \citenamefont
  {Lakra},\ and\ \citenamefont {Meriles}}]{Lozovoi2020}%
  \BibitemOpen
  \bibfield  {author} {\bibinfo {author} {\bibfnamefont {A.}~\bibnamefont
  {Lozovoi}}, \bibinfo {author} {\bibfnamefont {H.}~\bibnamefont {Jayakumar}},
  \bibinfo {author} {\bibfnamefont {D.}~\bibnamefont {Daw}}, \bibinfo {author}
  {\bibfnamefont {A.}~\bibnamefont {Lakra}},\ and\ \bibinfo {author}
  {\bibfnamefont {C.~A.}\ \bibnamefont {Meriles}},\ }\bibfield  {title}
  {\bibinfo {title} {Probing metastable space-charge potentials in a wide band
  gap semiconductor},\ }\href {https://doi.org/10.1103/PhysRevLett.125.256602}
  {\bibfield  {journal} {\bibinfo  {journal} {Phys. Rev. Lett.}\ }\textbf
  {\bibinfo {volume} {125}},\ \bibinfo {pages} {256602} (\bibinfo {year}
  {2020})}\BibitemShut {NoStop}%
\bibitem [{\citenamefont {Allers}\ and\ \citenamefont
  {Collins}(1995)}]{Collins1995}%
  \BibitemOpen
  \bibfield  {author} {\bibinfo {author} {\bibfnamefont {L.}~\bibnamefont
  {Allers}}\ and\ \bibinfo {author} {\bibfnamefont {A.~T.}\ \bibnamefont
  {Collins}},\ }\bibfield  {title} {\bibinfo {title} {Photoconductive
  spectroscopy of diamond grown by chemical vapor deposition},\ }\href
  {https://doi.org/10.1063/1.358566} {\bibfield  {journal} {\bibinfo  {journal}
  {Journal of Applied Physics}\ }\textbf {\bibinfo {volume} {77}},\ \bibinfo
  {pages} {3879} (\bibinfo {year} {1995})}\BibitemShut {NoStop}%
\bibitem [{\citenamefont {Nicolas}\ \emph {et~al.}(2019)\citenamefont
  {Nicolas}, \citenamefont {Delord}, \citenamefont {Huillery}, \citenamefont
  {Pellet-Mary},\ and\ \citenamefont {Hétet}}]{Louis2019}%
  \BibitemOpen
  \bibfield  {author} {\bibinfo {author} {\bibfnamefont {L.}~\bibnamefont
  {Nicolas}}, \bibinfo {author} {\bibfnamefont {T.}~\bibnamefont {Delord}},
  \bibinfo {author} {\bibfnamefont {P.}~\bibnamefont {Huillery}}, \bibinfo
  {author} {\bibfnamefont {C.}~\bibnamefont {Pellet-Mary}},\ and\ \bibinfo
  {author} {\bibfnamefont {G.}~\bibnamefont {Hétet}},\ }\bibfield  {title}
  {\bibinfo {title} {Sub-ghz linewidth ensembles of siv centers in a diamond
  nanopyramid revealed by charge state conversion},\ }\href
  {https://doi.org/10.1021/acsphotonics.9b00262} {\bibfield  {journal}
  {\bibinfo  {journal} {ACS Photonics}\ }\textbf {\bibinfo {volume} {6}},\
  \bibinfo {pages} {2413} (\bibinfo {year} {2019})}\BibitemShut {NoStop}%
\bibitem [{\citenamefont {Thiering}\ and\ \citenamefont
  {Gali}(2018)}]{Gergo2018}%
  \BibitemOpen
  \bibfield  {author} {\bibinfo {author} {\bibfnamefont {G.}~\bibnamefont
  {Thiering}}\ and\ \bibinfo {author} {\bibfnamefont {A.}~\bibnamefont
  {Gali}},\ }\bibfield  {title} {\bibinfo {title} {Ab initio magneto-optical
  spectrum of group-iv vacancy color centers in diamond},\ }\href
  {https://doi.org/10.1103/PhysRevX.8.021063} {\bibfield  {journal} {\bibinfo
  {journal} {Phys. Rev. X}\ }\textbf {\bibinfo {volume} {8}},\ \bibinfo {pages}
  {021063} (\bibinfo {year} {2018})}\BibitemShut {NoStop}%
\bibitem [{\citenamefont {Thiering}\ and\ \citenamefont
  {Gali}(2019)}]{Thiering2019}%
  \BibitemOpen
  \bibfield  {author} {\bibinfo {author} {\bibfnamefont {G.}~\bibnamefont
  {Thiering}}\ and\ \bibinfo {author} {\bibfnamefont {A.}~\bibnamefont
  {Gali}},\ }\bibfield  {title} {\bibinfo {title} {{The $(e_g \otimes e_u)
  \otimes E_g$ product Jahn–Teller effect in the neutral group-IV vacancy
  quantum bits in diamond}},\ }\href
  {https://doi.org/10.1038/s41524-019-0158-3} {\bibfield  {journal} {\bibinfo
  {journal} {npj Computational Materials}\ }\textbf {\bibinfo {volume} {5}},\
  \bibinfo {pages} {18} (\bibinfo {year} {2019})}\BibitemShut {NoStop}%
\end{thebibliography}%


\providecommand{\noopsort}[1]{}\providecommand{\singleletter}[1]{#1}%
\begin{thebibliography}{17}%
\makeatletter
\providecommand \@ifxundefined [1]{%
 \@ifx{#1\undefined}
}%
\providecommand \@ifnum [1]{%
 \ifnum #1\expandafter \@firstoftwo
 \else \expandafter \@secondoftwo
 \fi
}%
\providecommand \@ifx [1]{%
 \ifx #1\expandafter \@firstoftwo
 \else \expandafter \@secondoftwo
 \fi
}%
\providecommand \natexlab [1]{#1}%
\providecommand \enquote  [1]{``#1''}%
\providecommand \bibnamefont  [1]{#1}%
\providecommand \bibfnamefont [1]{#1}%
\providecommand \citenamefont [1]{#1}%
\providecommand \href@noop [0]{\@secondoftwo}%
\providecommand \href [0]{\begingroup \@sanitize@url \@href}%
\providecommand \@href[1]{\@@startlink{#1}\@@href}%
\providecommand \@@href[1]{\endgroup#1\@@endlink}%
\providecommand \@sanitize@url [0]{\catcode `\\12\catcode `\$12\catcode
  `\&12\catcode `\#12\catcode `\^12\catcode `\_12\catcode `\%12\relax}%
\providecommand \@@startlink[1]{}%
\providecommand \@@endlink[0]{}%
\providecommand \url  [0]{\begingroup\@sanitize@url \@url }%
\providecommand \@url [1]{\endgroup\@href {#1}{\urlprefix }}%
\providecommand \urlprefix  [0]{URL }%
\providecommand \Eprint [0]{\href }%
\providecommand \doibase [0]{https://doi.org/}%
\providecommand \selectlanguage [0]{\@gobble}%
\providecommand \bibinfo  [0]{\@secondoftwo}%
\providecommand \bibfield  [0]{\@secondoftwo}%
\providecommand \translation [1]{[#1]}%
\providecommand \BibitemOpen [0]{}%
\providecommand \bibitemStop [0]{}%
\providecommand \bibitemNoStop [0]{.\EOS\space}%
\providecommand \EOS [0]{\spacefactor3000\relax}%
\providecommand \BibitemShut  [1]{\csname bibitem#1\endcsname}%
\let\auto@bib@innerbib\@empty
\bibitem [{\citenamefont {Zhang}\ \emph {et~al.}(2015)\citenamefont {Zhang},
  \citenamefont {Braverman}, \citenamefont {Kawasaki},\ and\ \citenamefont
  {Vuletić}}]{Shutter2015}%
  \BibitemOpen
  \bibfield  {author} {\bibinfo {author} {\bibfnamefont {G.~H.}\ \bibnamefont
  {Zhang}}, \bibinfo {author} {\bibfnamefont {B.}~\bibnamefont {Braverman}},
  \bibinfo {author} {\bibfnamefont {A.}~\bibnamefont {Kawasaki}},\ and\
  \bibinfo {author} {\bibfnamefont {V.}~\bibnamefont {Vuletić}},\ }\href
  {https://doi.org/10.1063/1.4937614} {\bibfield  {journal} {\bibinfo
  {journal} {Review of Scientific Instruments}\ }\textbf {\bibinfo {volume}
  {86}},\ \bibinfo {pages} {126105} (\bibinfo {year} {2015})}\BibitemShut
  {NoStop}%
\bibitem [{\citenamefont {Zhang}\ \emph {et~al.}(2020)\citenamefont {Zhang},
  \citenamefont {Stevenson}, \citenamefont {Thiering}, \citenamefont {Rose},
  \citenamefont {Huang}, \citenamefont {Edmonds}, \citenamefont {Markham},
  \citenamefont {Lyon}, \citenamefont {Gali},\ and\ \citenamefont
  {de~Leon}}]{Zhang2020}%
  \BibitemOpen
  \bibfield  {author} {\bibinfo {author} {\bibfnamefont {Z.-H.}\ \bibnamefont
  {Zhang}}, \bibinfo {author} {\bibfnamefont {P.}~\bibnamefont {Stevenson}},
  \bibinfo {author} {\bibfnamefont {G.~m.~H.}\ \bibnamefont {Thiering}},
  \bibinfo {author} {\bibfnamefont {B.~C.}\ \bibnamefont {Rose}}, \bibinfo
  {author} {\bibfnamefont {D.}~\bibnamefont {Huang}}, \bibinfo {author}
  {\bibfnamefont {A.~M.}\ \bibnamefont {Edmonds}}, \bibinfo {author}
  {\bibfnamefont {M.~L.}\ \bibnamefont {Markham}}, \bibinfo {author}
  {\bibfnamefont {S.~A.}\ \bibnamefont {Lyon}}, \bibinfo {author}
  {\bibfnamefont {A.}~\bibnamefont {Gali}},\ and\ \bibinfo {author}
  {\bibfnamefont {N.~P.}\ \bibnamefont {de~Leon}},\ }\href
  {https://doi.org/10.1103/PhysRevLett.125.237402} {\bibfield  {journal}
  {\bibinfo  {journal} {Phys. Rev. Lett.}\ }\textbf {\bibinfo {volume} {125}},\
  \bibinfo {pages} {237402} (\bibinfo {year} {2020})}\BibitemShut {NoStop}%
\bibitem [{\citenamefont {Rose}\ \emph {et~al.}(2017)\citenamefont {Rose},
  \citenamefont {Weis}, \citenamefont {Tyryshkin}, \citenamefont {Schenkel},\
  and\ \citenamefont {Lyon}}]{Rose2017}%
  \BibitemOpen
  \bibfield  {author} {\bibinfo {author} {\bibfnamefont {B.}~\bibnamefont
  {Rose}}, \bibinfo {author} {\bibfnamefont {C.}~\bibnamefont {Weis}}, \bibinfo
  {author} {\bibfnamefont {A.}~\bibnamefont {Tyryshkin}}, \bibinfo {author}
  {\bibfnamefont {T.}~\bibnamefont {Schenkel}},\ and\ \bibinfo {author}
  {\bibfnamefont {S.}~\bibnamefont {Lyon}},\ }\href
  {https://doi.org/https://doi.org/10.1016/j.diamond.2016.12.009} {\bibfield
  {journal} {\bibinfo  {journal} {Diamond and Related Materials}\ }\textbf
  {\bibinfo {volume} {72}},\ \bibinfo {pages} {32} (\bibinfo {year}
  {2017})}\BibitemShut {NoStop}%
\bibitem [{\citenamefont {Rose}\ \emph {et~al.}(2018)\citenamefont {Rose},
  \citenamefont {Huang}, \citenamefont {Zhang}, \citenamefont {Stevenson},
  \citenamefont {Tyryshkin}, \citenamefont {Sangtawesin}, \citenamefont
  {Srinivasan}, \citenamefont {Loudin}, \citenamefont {Markham}, \citenamefont
  {Edmonds}, \citenamefont {Twitchen}, \citenamefont {Lyon},\ and\
  \citenamefont {de~Leon}}]{Rose2018}%
  \BibitemOpen
  \bibfield  {author} {\bibinfo {author} {\bibfnamefont {B.~C.}\ \bibnamefont
  {Rose}}, \bibinfo {author} {\bibfnamefont {D.}~\bibnamefont {Huang}},
  \bibinfo {author} {\bibfnamefont {Z.-H.}\ \bibnamefont {Zhang}}, \bibinfo
  {author} {\bibfnamefont {P.}~\bibnamefont {Stevenson}}, \bibinfo {author}
  {\bibfnamefont {A.~M.}\ \bibnamefont {Tyryshkin}}, \bibinfo {author}
  {\bibfnamefont {S.}~\bibnamefont {Sangtawesin}}, \bibinfo {author}
  {\bibfnamefont {S.}~\bibnamefont {Srinivasan}}, \bibinfo {author}
  {\bibfnamefont {L.}~\bibnamefont {Loudin}}, \bibinfo {author} {\bibfnamefont
  {M.~L.}\ \bibnamefont {Markham}}, \bibinfo {author} {\bibfnamefont {A.~M.}\
  \bibnamefont {Edmonds}}, \bibinfo {author} {\bibfnamefont {D.~J.}\
  \bibnamefont {Twitchen}}, \bibinfo {author} {\bibfnamefont {S.~A.}\
  \bibnamefont {Lyon}},\ and\ \bibinfo {author} {\bibfnamefont {N.~P.}\
  \bibnamefont {de~Leon}},\ }\href {https://doi.org/10.1126/science.aao0290}
  {\bibfield  {journal} {\bibinfo  {journal} {Science}\ }\textbf {\bibinfo
  {volume} {361}},\ \bibinfo {pages} {60} (\bibinfo {year} {2018})}\BibitemShut
  {NoStop}%
\bibitem [{\citenamefont {Liggins}(2010)}]{Liggins2010}%
  \BibitemOpen
  \bibfield  {author} {\bibinfo {author} {\bibfnamefont {S.}~\bibnamefont
  {Liggins}},\ }\href {http://webcat.warwick.ac.uk/record=b2491628$\sim$S15}
  {\bibfield  {journal} {\bibinfo  {journal} {Th{\`{e}}se}\ } (\bibinfo {year}
  {2010})}\BibitemShut {NoStop}%
\bibitem [{\citenamefont {Edmonds}\ \emph {et~al.}(2012)\citenamefont
  {Edmonds}, \citenamefont {D'Haenens-Johansson}, \citenamefont {Cruddace},
  \citenamefont {Newton}, \citenamefont {Fu}, \citenamefont {Santori},
  \citenamefont {Beausoleil}, \citenamefont {Twitchen},\ and\ \citenamefont
  {Markham}}]{Edmonds2012}%
  \BibitemOpen
  \bibfield  {author} {\bibinfo {author} {\bibfnamefont {A.~M.}\ \bibnamefont
  {Edmonds}}, \bibinfo {author} {\bibfnamefont {U.~F.~S.}\ \bibnamefont
  {D'Haenens-Johansson}}, \bibinfo {author} {\bibfnamefont {R.~J.}\
  \bibnamefont {Cruddace}}, \bibinfo {author} {\bibfnamefont {M.~E.}\
  \bibnamefont {Newton}}, \bibinfo {author} {\bibfnamefont {K.-M.~C.}\
  \bibnamefont {Fu}}, \bibinfo {author} {\bibfnamefont {C.}~\bibnamefont
  {Santori}}, \bibinfo {author} {\bibfnamefont {R.~G.}\ \bibnamefont
  {Beausoleil}}, \bibinfo {author} {\bibfnamefont {D.~J.}\ \bibnamefont
  {Twitchen}},\ and\ \bibinfo {author} {\bibfnamefont {M.~L.}\ \bibnamefont
  {Markham}},\ }\href {https://doi.org/10.1103/PhysRevB.86.035201} {\bibfield
  {journal} {\bibinfo  {journal} {Phys. Rev. B}\ }\textbf {\bibinfo {volume}
  {86}},\ \bibinfo {pages} {035201} (\bibinfo {year} {2012})}\BibitemShut
  {NoStop}%
\bibitem [{\citenamefont {Jayakumar}\ \emph {et~al.}(2016)\citenamefont
  {Jayakumar}, \citenamefont {Henshaw}, \citenamefont {Dhomkar}, \citenamefont
  {Pagliero}, \citenamefont {Laraoui}, \citenamefont {Manson}, \citenamefont
  {Albu}, \citenamefont {Doherty},\ and\ \citenamefont
  {Meriles}}]{Jayakumar2016}%
  \BibitemOpen
  \bibfield  {author} {\bibinfo {author} {\bibfnamefont {H.}~\bibnamefont
  {Jayakumar}}, \bibinfo {author} {\bibfnamefont {J.}~\bibnamefont {Henshaw}},
  \bibinfo {author} {\bibfnamefont {S.}~\bibnamefont {Dhomkar}}, \bibinfo
  {author} {\bibfnamefont {D.}~\bibnamefont {Pagliero}}, \bibinfo {author}
  {\bibfnamefont {A.}~\bibnamefont {Laraoui}}, \bibinfo {author} {\bibfnamefont
  {N.~B.}\ \bibnamefont {Manson}}, \bibinfo {author} {\bibfnamefont
  {R.}~\bibnamefont {Albu}}, \bibinfo {author} {\bibfnamefont {M.~W.}\
  \bibnamefont {Doherty}},\ and\ \bibinfo {author} {\bibfnamefont {C.~A.}\
  \bibnamefont {Meriles}},\ }\href {https://doi.org/10.1038/ncomms12660}
  {\bibfield  {journal} {\bibinfo  {journal} {Nature Communications}\ }\textbf
  {\bibinfo {volume} {7}},\ \bibinfo {pages} {12660} (\bibinfo {year}
  {2016})}\BibitemShut {NoStop}%
\bibitem [{\citenamefont {Dhomkar}\ \emph {et~al.}(2018)\citenamefont
  {Dhomkar}, \citenamefont {Zangara}, \citenamefont {Henshaw},\ and\
  \citenamefont {Meriles}}]{Dhomkar2018}%
  \BibitemOpen
  \bibfield  {author} {\bibinfo {author} {\bibfnamefont {S.}~\bibnamefont
  {Dhomkar}}, \bibinfo {author} {\bibfnamefont {P.~R.}\ \bibnamefont
  {Zangara}}, \bibinfo {author} {\bibfnamefont {J.}~\bibnamefont {Henshaw}},\
  and\ \bibinfo {author} {\bibfnamefont {C.~A.}\ \bibnamefont {Meriles}},\
  }\href {https://doi.org/10.1103/PhysRevLett.120.117401} {\bibfield  {journal}
  {\bibinfo  {journal} {Phys. Rev. Lett.}\ }\textbf {\bibinfo {volume} {120}},\
  \bibinfo {pages} {117401} (\bibinfo {year} {2018})}\BibitemShut {NoStop}%
\bibitem [{\citenamefont {Lozovoi}\ \emph {et~al.}(2020)\citenamefont
  {Lozovoi}, \citenamefont {Jayakumar}, \citenamefont {Daw}, \citenamefont
  {Lakra},\ and\ \citenamefont {Meriles}}]{Lozovoi2020}%
  \BibitemOpen
  \bibfield  {author} {\bibinfo {author} {\bibfnamefont {A.}~\bibnamefont
  {Lozovoi}}, \bibinfo {author} {\bibfnamefont {H.}~\bibnamefont {Jayakumar}},
  \bibinfo {author} {\bibfnamefont {D.}~\bibnamefont {Daw}}, \bibinfo {author}
  {\bibfnamefont {A.}~\bibnamefont {Lakra}},\ and\ \bibinfo {author}
  {\bibfnamefont {C.~A.}\ \bibnamefont {Meriles}},\ }\href
  {https://doi.org/10.1103/PhysRevLett.125.256602} {\bibfield  {journal}
  {\bibinfo  {journal} {Phys. Rev. Lett.}\ }\textbf {\bibinfo {volume} {125}},\
  \bibinfo {pages} {256602} (\bibinfo {year} {2020})}\BibitemShut {NoStop}%
\bibitem [{\citenamefont {Allers}\ and\ \citenamefont
  {Collins}(1995)}]{Collins1995}%
  \BibitemOpen
  \bibfield  {author} {\bibinfo {author} {\bibfnamefont {L.}~\bibnamefont
  {Allers}}\ and\ \bibinfo {author} {\bibfnamefont {A.~T.}\ \bibnamefont
  {Collins}},\ }\href {https://doi.org/10.1063/1.358566} {\bibfield  {journal}
  {\bibinfo  {journal} {Journal of Applied Physics}\ }\textbf {\bibinfo
  {volume} {77}},\ \bibinfo {pages} {3879} (\bibinfo {year}
  {1995})}\BibitemShut {NoStop}%
\bibitem [{\citenamefont {Lozovoi}\ \emph {et~al.}(2021)\citenamefont
  {Lozovoi}, \citenamefont {Jayakumar}, \citenamefont {Daw}, \citenamefont
  {Vizkelethy}, \citenamefont {Bielejec}, \citenamefont {Doherty},
  \citenamefont {Flick},\ and\ \citenamefont {Meriles}}]{Lozovoi2021}%
  \BibitemOpen
  \bibfield  {author} {\bibinfo {author} {\bibfnamefont {A.}~\bibnamefont
  {Lozovoi}}, \bibinfo {author} {\bibfnamefont {H.}~\bibnamefont {Jayakumar}},
  \bibinfo {author} {\bibfnamefont {D.}~\bibnamefont {Daw}}, \bibinfo {author}
  {\bibfnamefont {G.}~\bibnamefont {Vizkelethy}}, \bibinfo {author}
  {\bibfnamefont {E.}~\bibnamefont {Bielejec}}, \bibinfo {author}
  {\bibfnamefont {M.~W.}\ \bibnamefont {Doherty}}, \bibinfo {author}
  {\bibfnamefont {J.}~\bibnamefont {Flick}},\ and\ \bibinfo {author}
  {\bibfnamefont {C.~A.}\ \bibnamefont {Meriles}},\ }\href
  {https://doi.org/10.1038/s41928-021-00656-z} {\bibfield  {journal} {\bibinfo
  {journal} {Nature Electronics}\ }\textbf {\bibinfo {volume} {4}},\ \bibinfo
  {pages} {717} (\bibinfo {year} {2021})}\BibitemShut {NoStop}%
\bibitem [{\citenamefont {Farrer}(1969)}]{Farrer1969}%
  \BibitemOpen
  \bibfield  {author} {\bibinfo {author} {\bibfnamefont {R.}~\bibnamefont
  {Farrer}},\ }\href
  {https://doi.org/https://doi.org/10.1016/0038-1098(69)90593-6} {\bibfield
  {journal} {\bibinfo  {journal} {Solid State Communications}\ }\textbf
  {\bibinfo {volume} {7}},\ \bibinfo {pages} {685} (\bibinfo {year}
  {1969})}\BibitemShut {NoStop}%
\bibitem [{\citenamefont {Nesl{\'{a}}dek}\ \emph {et~al.}(1998)\citenamefont
  {Nesl{\'{a}}dek}, \citenamefont {Stals}, \citenamefont {Stesmans},
  \citenamefont {Iakoubovskij}, \citenamefont {Adriaenssens}, \citenamefont
  {Rosa},\ and\ \citenamefont {Van{\v{e}}{\v{c}}ek}}]{Nesladek1998}%
  \BibitemOpen
  \bibfield  {author} {\bibinfo {author} {\bibfnamefont {M.}~\bibnamefont
  {Nesl{\'{a}}dek}}, \bibinfo {author} {\bibfnamefont {L.~M.}\ \bibnamefont
  {Stals}}, \bibinfo {author} {\bibfnamefont {A.}~\bibnamefont {Stesmans}},
  \bibinfo {author} {\bibfnamefont {K.}~\bibnamefont {Iakoubovskij}}, \bibinfo
  {author} {\bibfnamefont {G.~J.}\ \bibnamefont {Adriaenssens}}, \bibinfo
  {author} {\bibfnamefont {J.}~\bibnamefont {Rosa}},\ and\ \bibinfo {author}
  {\bibfnamefont {M.}~\bibnamefont {Van{\v{e}}{\v{c}}ek}},\ }\href
  {https://doi.org/10.1063/1.121632} {\bibfield  {journal} {\bibinfo  {journal}
  {Appl. Phys. Lett.}\ }\textbf {\bibinfo {volume} {72}},\ \bibinfo {pages}
  {3306} (\bibinfo {year} {1998})}\BibitemShut {NoStop}%
\bibitem [{\citenamefont {Isberg}\ \emph {et~al.}(2006)\citenamefont {Isberg},
  \citenamefont {Tajani},\ and\ \citenamefont {Twitchen}}]{Isberg2006}%
  \BibitemOpen
  \bibfield  {author} {\bibinfo {author} {\bibfnamefont {J.}~\bibnamefont
  {Isberg}}, \bibinfo {author} {\bibfnamefont {A.}~\bibnamefont {Tajani}},\
  and\ \bibinfo {author} {\bibfnamefont {D.~J.}\ \bibnamefont {Twitchen}},\
  }\href {https://doi.org/10.1103/PhysRevB.73.245207} {\bibfield  {journal}
  {\bibinfo  {journal} {Phys. Rev. B - Condens. Matter Mater. Phys.}\ }\textbf
  {\bibinfo {volume} {73}},\ \bibinfo {pages} {1} (\bibinfo {year}
  {2006})}\BibitemShut {NoStop}%
\bibitem [{\citenamefont {Heremans}\ \emph {et~al.}(2009)\citenamefont
  {Heremans}, \citenamefont {Fuchs}, \citenamefont {Wang}, \citenamefont
  {Hanson},\ and\ \citenamefont {Awschalom}}]{Heremans2009}%
  \BibitemOpen
  \bibfield  {author} {\bibinfo {author} {\bibfnamefont {F.~J.}\ \bibnamefont
  {Heremans}}, \bibinfo {author} {\bibfnamefont {G.~D.}\ \bibnamefont {Fuchs}},
  \bibinfo {author} {\bibfnamefont {C.~F.}\ \bibnamefont {Wang}}, \bibinfo
  {author} {\bibfnamefont {R.}~\bibnamefont {Hanson}},\ and\ \bibinfo {author}
  {\bibfnamefont {D.~D.}\ \bibnamefont {Awschalom}},\ }\href
  {https://doi.org/10.1063/1.3120225} {\bibfield  {journal} {\bibinfo
  {journal} {Applied Physics Letters}\ }\textbf {\bibinfo {volume} {94}},\
  \bibinfo {pages} {152102} (\bibinfo {year} {2009})}\BibitemShut {NoStop}%
\bibitem [{\citenamefont {Gardill}\ \emph {et~al.}(2021)\citenamefont
  {Gardill}, \citenamefont {Kemeny}, \citenamefont {Cambria}, \citenamefont
  {Li}, \citenamefont {Dinani}, \citenamefont {Norambuena}, \citenamefont
  {Maze}, \citenamefont {Lordi},\ and\ \citenamefont
  {Kolkowitz}}]{Gardill2021}%
  \BibitemOpen
  \bibfield  {author} {\bibinfo {author} {\bibfnamefont {A.}~\bibnamefont
  {Gardill}}, \bibinfo {author} {\bibfnamefont {I.}~\bibnamefont {Kemeny}},
  \bibinfo {author} {\bibfnamefont {M.~C.}\ \bibnamefont {Cambria}}, \bibinfo
  {author} {\bibfnamefont {Y.}~\bibnamefont {Li}}, \bibinfo {author}
  {\bibfnamefont {H.~T.}\ \bibnamefont {Dinani}}, \bibinfo {author}
  {\bibfnamefont {A.}~\bibnamefont {Norambuena}}, \bibinfo {author}
  {\bibfnamefont {J.~R.}\ \bibnamefont {Maze}}, \bibinfo {author}
  {\bibfnamefont {V.}~\bibnamefont {Lordi}},\ and\ \bibinfo {author}
  {\bibfnamefont {S.}~\bibnamefont {Kolkowitz}},\ }\href
  {https://doi.org/10.1021/acs.nanolett.1c02250} {\bibfield  {journal}
  {\bibinfo  {journal} {Nano Letters}\ }\textbf {\bibinfo {volume} {21}},\
  \bibinfo {pages} {6960} (\bibinfo {year} {2021})}\BibitemShut {NoStop}%
\bibitem [{\citenamefont {Gali}\ and\ \citenamefont {Maze}(2013)}]{Gali2013}%
  \BibitemOpen
  \bibfield  {author} {\bibinfo {author} {\bibfnamefont {A.}~\bibnamefont
  {Gali}}\ and\ \bibinfo {author} {\bibfnamefont {J.~R.}\ \bibnamefont
  {Maze}},\ }\href {https://doi.org/10.1103/PhysRevB.88.235205} {\bibfield
  {journal} {\bibinfo  {journal} {Phys. Rev. B}\ }\textbf {\bibinfo {volume}
  {88}},\ \bibinfo {pages} {235205} (\bibinfo {year} {2013})}\BibitemShut
  {NoStop}%
\end{thebibliography}%

\end{document}


\title{Supplemental Material for \\``Neutral Silicon Vacancy Centers in Diamond via Photoactivated Itinerant Carriers''}
\author{Zi-Huai Zhang}
\affiliation{%
Department of Electrical and Computer Engineering, Princeton University, Princeton, New Jersey 08544, USA
}%

\author{Andrew M.
Edmonds}
\affiliation{
Element Six, Harwell, OX11 0QR, United Kingdom
}

\author{Nicola Palmer}
\affiliation{
Element Six, Harwell, OX11 0QR, United Kingdom
}

\author{Matthew L. Markham}
\affiliation{
Element Six, Harwell, OX11 0QR, United Kingdom
}

\author{Nathalie P. de Leon}%
 \email{npdeleon@princeton.edu}
\affiliation{%
Department of Electrical and Computer Engineering, Princeton University, Princeton, New Jersey 08544, USA
}%

\date{\today}
\maketitle
\label{Sec:SI}

\setcounter{figure}{0}
\setcounter{section}{0}
\renewcommand{\thefigure}{S\arabic{figure}}
\renewcommand{\thetable}{\Roman{table}}
\renewcommand{\thesection}{\Roman{section}}

\section{\label{SI_Setup} SUPPLEMENTARY EXPERIMENTAL METHODS}
\subsection{Experimental Setup}
Optical measurements were performed in a home-built confocal cryogenic confocal microscope with a helium flow cryostat (Janis ST-500 probe station). The confocal microscope contains two independent branches for excitation and detection for  near-infrared (NIR) and visible wavelengths. Both branches are equipped with a scanning galvo system (Thorlabs GVS012). The two branches are combined with a pellicle beamsplitter (Thorlabs BP245B1) and directed into a 4f system. Finally, a NIR 50X objective lens (Olympus LCPLN50XIR) inside vacuum is used to focus excitation onto the sample and collect the fluorescence signal.

The NIR branch of the confocal microscope was used for neutral silicon vacancy (\siv{}) measurement. The excitation channel and detection channel are combined with a 925 nm dichroic beamsplitter (Semrock FF925-Di01-25-D). The resonant excitation channel is combined with the detection channel with a 10/90 beamspliter (Thorlabs BS044). Off-resonant excitation is performed with a tunable diode laser (Toptica DL pro 850 nm), and the signal is filtered using a 937 nm long pass filter (Semrock FF01-937/LP-25). The laser is pulsed using a home-built shutter system \cite{Shutter2015}, and intensity controlled by a variable optical attenuator (Thorlabs V800PA). Photoluminescence excitation spectroscopy at the zero-phonon line is performed using resonant excitation with a tunable diode laser (Toptica CTL 950) and detecting the sideband emission of \siv{} with a 980~nm long pass filter (Semrock LP02-980RE-25). The signal is coupled into a single mode fiber and detected either by a CCD spectrometer (Princeton Instruments Acton SP-2300i with Pixis 100 CCD and 300 g/mm grating) or by a superconducting nanowire detector (Quantum Opus, optimized for 950 - 1100 nm). For ODMR, microwave (MW) excitation is applied using a thin wire stretched across the sample. The MW excitation is generated with a signal generator (Keysight N9310A) and then amplified by a high-power MW amplifier (Triad TB1003). Two 0.8 - 2 GHz MW circulators (Ditom D3C0802S) were added after the amplifier for circuit protection. The MW excitation is pulsed using a fast MW switch (Mini-Circuits ZASWA-2-50DR+) gated by a TTL pulse generator (Spincore PBESR-PRO-500-PCI). For ODMR, MW excitation is modulated to have a 2 ms period with 50\% duty cycle.

The visible branch of the confocal microscope was used for measurements of negatively charged silicon vacancy centers (\sivm{}) and for visible wavelength photodoping to form  \siv{}.  The excitation and detection channels are combined with a 650 nm dichroic beamsplitter (Thorlabs DMLP650). The microscope is equipped with three different lasers for 532 nm (Lamdapro UG-100 mW), 594 nm (Newport R-39582), and 637 nm (Thorlabs LP637-SF70) excitation. The three lasers are coupled to a single fiber using a RGB Combiner (Thorlabs RGB26HF). The 637 nm laser is pulsed using a home-built shutter system \cite{Shutter2015}. The 532 nm laser is pulsed using an acousto-optic modulator and intensity controlled by a variable optical attenuator (Thorlabs V600A). For fluorescence measurement of \siv{}, the signal is filtered by a bandpass filter (Thorlabs FB740-10) and detected by a single photon detector (Excelitas SPCM-AQRH-44-FC).

\subsection{Sample Preparation}
A \{110\} diamond (D1) grown by plasma chemical-vapor deposition (Element Six) was studied in the main text. The diamond was doped with silicon during growth, and the silicon concentration in the sample is measured to be 0.8~ppm based on secondary ion mass spectrometry. The concentration of \sivm{} centers is estimated to be $\sim$30 ppb by comparing the \sivm{} fluorescence with a sample of known \sivm{} concentration. This sample contains \siv{} centers in some regions, and the optical spin polarization of these \siv{} centers was previously studied in Ref.~\cite{Zhang2020}. Throughout this work, we work in regions where no \siv{} can be observed without photoactivated itinerant carriers. The concentration of the nitrogen vacancy (NV$^-$) centers is estimated to be $\sim$0.03 ppb by comparing the NV$^-$ fluorescence with a sample of known NV$^-$ concentration~\cite{Rose2017}.

Data based on a second silicon doped diamond (D2) grown by plasma chemical-vapor deposition (Element Six) is presented in the supplemental material. The concentration of \sivm{} centers is estimated to be $\sim$40 ppb using UV-Vis absorption. The concentration of NV$^-$ centers in this sample is estimated to be $\sim$0.07 ppb. In this sample, we observe stabilization of nonequilibrium \siv{}, and similar wavelength dependent photoactivation of itinerant carriers.

Data based on a third silicon doped diamond (D3) grown by plasma chemical-vapor deposition (Element Six) is presented in the supplemental material to demonstrate the heterogeneous charge dynamics behavior among different samples and the spectroscopic signature of SiV$^{2-}$. The sample is doped with nitrogen and silicon during growth, with an estimated nitrogen concentration of 50 ppb based on growth conditions and a silicon concentration of 300 ppb based on secondary ion mass spectrometry (SIMS) measurement. We estimate the resulting NV$^-$ concentration to be $\sim$0.01 ppb and the \sivm{} concentration to be $\sim$2 ppb by comparing the NV$^-$ fluorescence and the \sivm{} fluorescence to fluorescence in samples with known NV$^-$ and \sivm{} concentration.

Data based on a fourth diamond (D4) grown by plasma chemical-vapor deposition (Element Six) is presented in the supplemental material for study of single center dynamics. The optical properties of the single \siv{} centers in this sample were previously characterized in Ref.~\cite{Rose2018}. The concentration of NV$^-$ centers in this sample is not uniform, and is estimated to be in the range of 0.1 - 0.9 ppb. This sample was implanted with silicon with a fluence of $1\times 10^9$~cm$^{-2}$. After implantation, the sample was annealed up to 1200 $\degree$C with the following steps: (1) Ramp to 100 $^\circ$C over 1 hour, hold for 11 hours; (2) Ramp to 400 $^\circ$C over 4 hours, hold for 8 hours; (3) Ramp to 800 $^\circ$C over 6 hours, hold for 8 hours; (4) Ramp to 1200 $^\circ$C over 6 hours, hold for 2 hours; (4) Let cool to room temperature. Single centers are observable after the annealing. 

\begin{figure}[h!]
  \centering
  \includegraphics[width = 129mm]{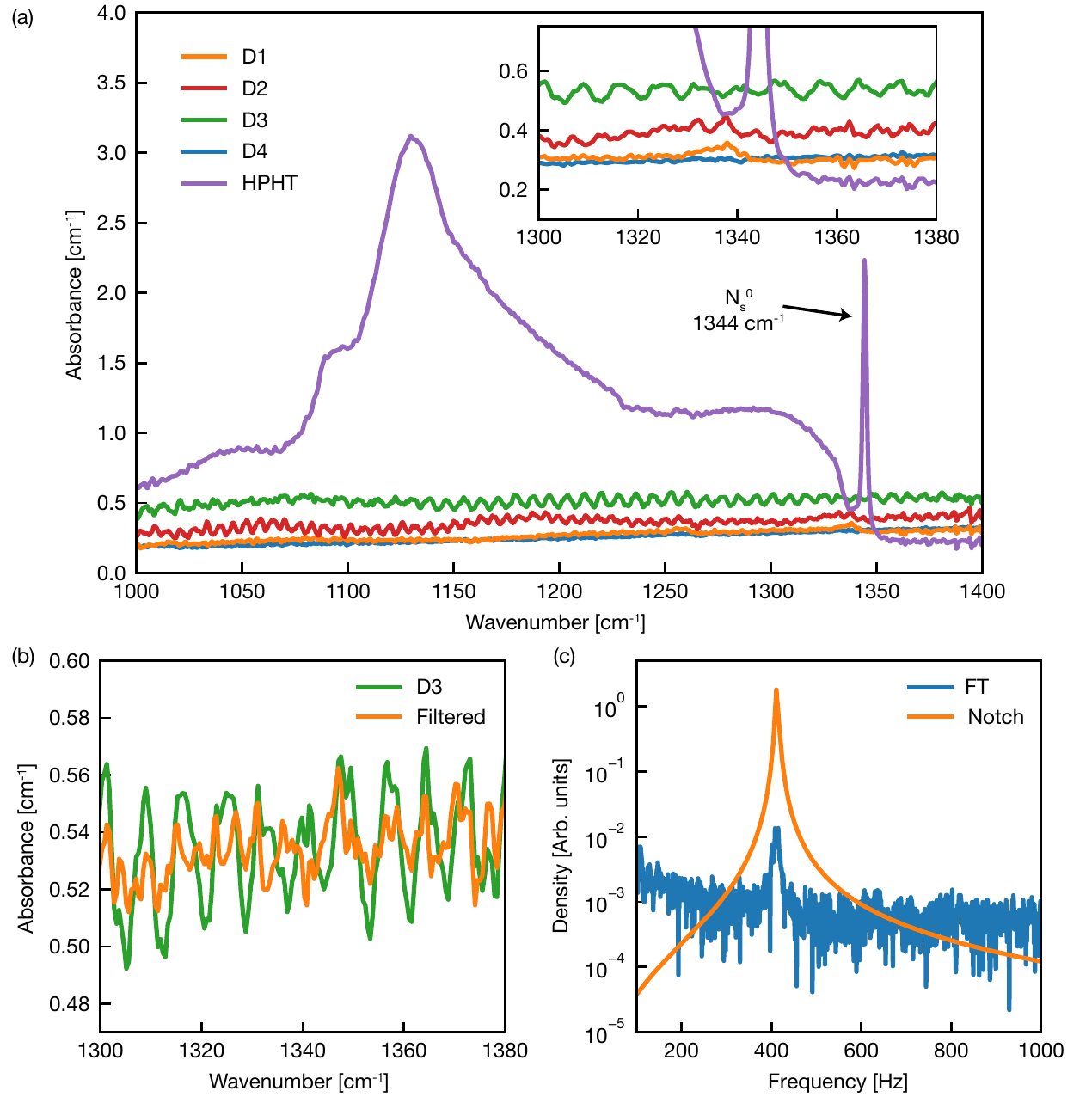}
  \caption{{\bf FTIR measurements}. (a) FTIR spectra on different diamond samples. A HPHT diamond with high nitrogen concentration is plotted for reference. The peak at 1344 cm$^{-1}$ is characteristic to $\mathrm{N}_s^0$ and the peak height can converted to the $\mathrm{N}_s^0$ concentration with a converstion factor of 30 ppm per cm$^{-1}$ absorbance. Inset: FTIR spectra near the expected $\mathrm{N}_s^0$ peak. No peak can be identified except for the HPHT diamond. (b) FTIR spectrum for sample D3 before and after the noise filtering. (c) Fourier transform of FTIR spectrum on sample D3 (blue). The interference induced oscillation leads to a peak at 412~Hz. A notch filter centerd at 412~Hz with a bandwidth of 10~Hz is used to filter out the background oscillation (yellow).} 
  \label{fig:FTIR}
\end{figure}

We use Fourier-transform infrared spectroscopy (FTIR) to estimate the nitrogen concentration in the samples. The 1344 cm$^{-1}$ peak in the FTIR spectrum is related to substitutional nitrogen ($\mathrm{N}_s^0$, also referred to as the P1 center) and can be used for quantitative estimation of the concentration \cite{Liggins2010} (Fig.~\ref{fig:FTIR}(a)). In some samples, background oscillations show up in the FTIR spectrum due to interference of reflections from the two surfaces. We mitigate the contribution from this periodic background by filtering the spectrum with a narrow-band notch filter (Fig.~\ref{fig:FTIR}(b) and Fig.~\ref{fig:FTIR}(c)). No peak from $\mathrm{N}_s^0$ can be observed. Based on the noise level (standard deviation) from 1360 cm$^{-1}$ to 1380 cm$^{-1}$, we put a conservative upper bound of 300 ppb for $\mathrm{N}_s^0$ concentration in these samples.

It was reported that the ratio between total nitrogen concentration and NV$^-$ concentration is typically around 300:1 in nitrogen doped CVD diamonds~\cite{Edmonds2012}. Using this conversion factor, the nitrogen concentration is projected to be 9 ppb, 21 ppb and 3 ppb for sample D1, D2 and D3 based on the estimated native NV$^-$ concentration. We note that the projected nitrogen concentration of 3 ppb for sample D3 is smaller than the estimation (50~ppb) from growth condition. This difference suggests that the incorporation of silicon likely influences the concentration ratio between nitrogen and NV$^-$. Therefore, the projected nitrogen concentration here will be a conservative lower bound.
\section{Additional Measurements}
\subsection{Charge state dynamics in the dark}
\begin{figure}[h!]
  \centering
  \includegraphics[width = 129mm]{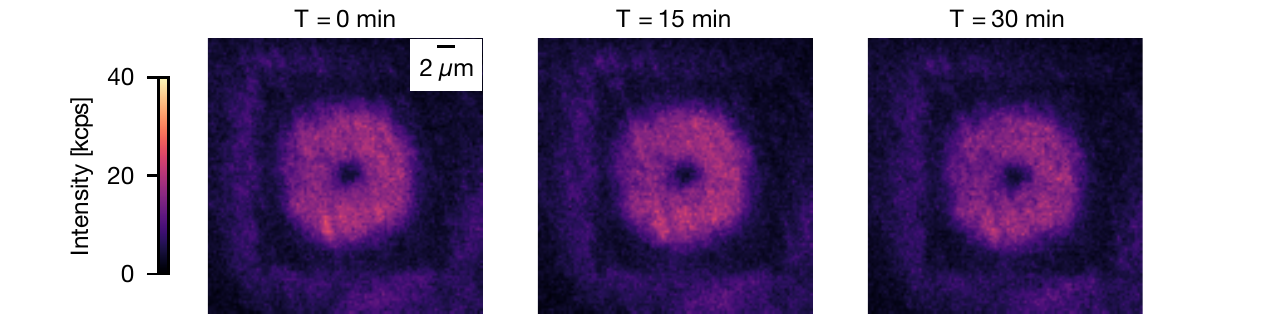}
  \caption{{\bf Charge state stability in the dark on sample D1}. After fixed location 532~nm illumination, nonequilibrium \siv{} centers are probed after dark periods of 15 minutes and 30 minutes with 0.28~mW of 857 nm. The spatial distribution of \siv{} remains stable, suggesting \siv{} centers generated in this nonequilibrium charge environment are long lived in the dark.}
  \label{fig:LA_dark}
\end{figure}

We probe the population decay of nonequilibrium \siv{} centers in the dark after they are generated. No decay can be observed after 30~min (Fig.~\ref{fig:LA_dark}), consistent with previous observation of long-lived nonequilibrium charge states for NV centers and \sivm{} centers \cite{Jayakumar2016,Dhomkar2018}.

\subsection{Time-dependent study of carrier transport}
In this section, we analyze the time-dependent carrier transport and capture by studying the hole diffusion process as a function of illumination time and excitation power.

\begin{figure}[h!]
  \centering
  \includegraphics[width = 129mm]{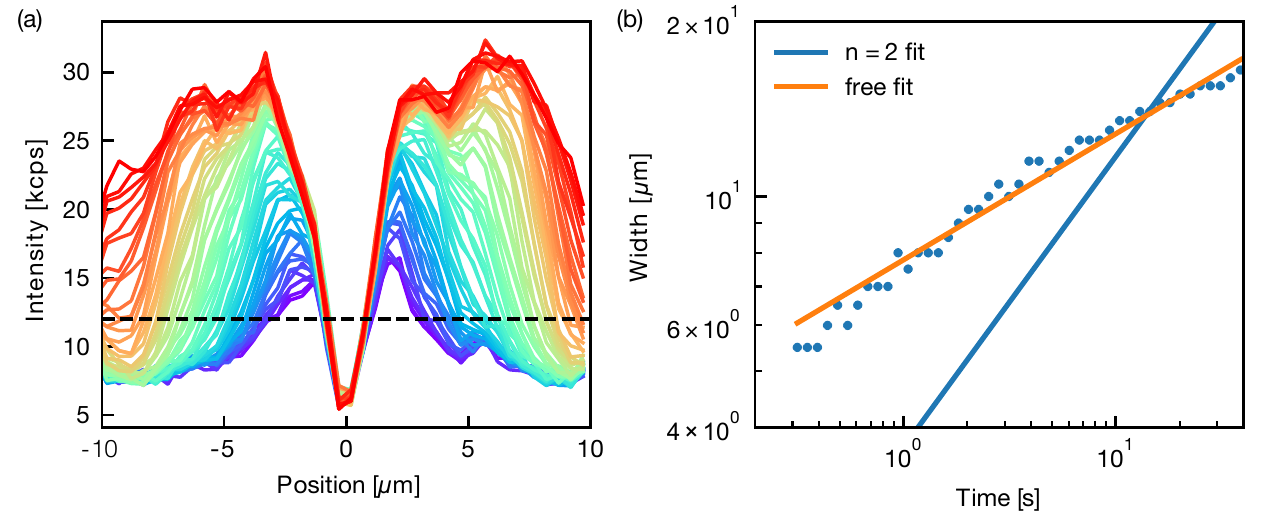}
  \caption{{\bf Time dependent measurement of \siv{} generation}. (a) Spatial distribution of \siv{} as a function of 532~nm (0.95~mW) illumination time. The duration was set from 0.3~s to 199.5~s with logarithmic spacing. The threshold for edge detection was set to 12 kcps. (b) Width of the \siv{} torus as a function of total 532~nm illumination time. The data was fitted with a model $\sigma(t) = D^{1/2}t^{1/n}$, where $t$ denotes the total 532~nm time, $D$ is a free coefficient, and n is the exponent describing the speed for the growth of the torus. The blue curve is a fit to the data with $n$ constrained to 2 while the yellow curve is a fit to the date with $n$ as a free parameter, with $n = 4.6 \pm 0.1$.}
  \label{fig:Diffusion_Time}
\end{figure}

The width of the \siv{} distribution is extracted by detecting the edge of the bright torus with a fixed threshold. Its evolution as a function of 532~nm illumination duration is shown in (Fig.~\ref{fig:Diffusion_Time}(a)). For a diffusion process described by Brownian motion, the width ($\sigma$) of the diffusion can be described by $\sigma(t) = (2D_{eff})^{1/2}t^{1/2}$ where $D_{eff}$ is the effective diffusion coefficient. We observe that the growth rate of the torus deviates significantly from the $t^{1/2}$ scaling with a fitted scaling of $t^{1/4.6}$. This slower growth rate suggests that the carrier diffusion cannot be described with the simple diffusion model. By inspecting the \siv{} profile as a function of time, we observe that the saturated \siv{} intensity reaches a similar value at different locations, and the farther away \sivm{} centers are converted to \siv{} only when the closer \sivm{} centers are already converted. This suggests that the \sivm{} centers are acting as a strong absorptive medium for the holes, limiting the rate of hole diffusion. 

To probe the hole diffusion dynamics before the saturation of \siv{} centers, we performed the same measurement with lower 532~nm power. Two different scalings in the growth rate of the torus are observed (Fig. S3(b)). At early times, the size of torus grows with a scaling close to $n = 2$, consistent with the diffusion model. In the late times where the saturation of \siv{} becomes more prominent, the growth rate of the torus slows down and follows a scaling of $n = 4.0 \pm 0.1$, similar to the scaling observed in the higher power measurements (Fig.~\ref{fig:Diffusion_Time}).

\begin{figure}[h!]
  \centering
  \includegraphics[width = 129mm]{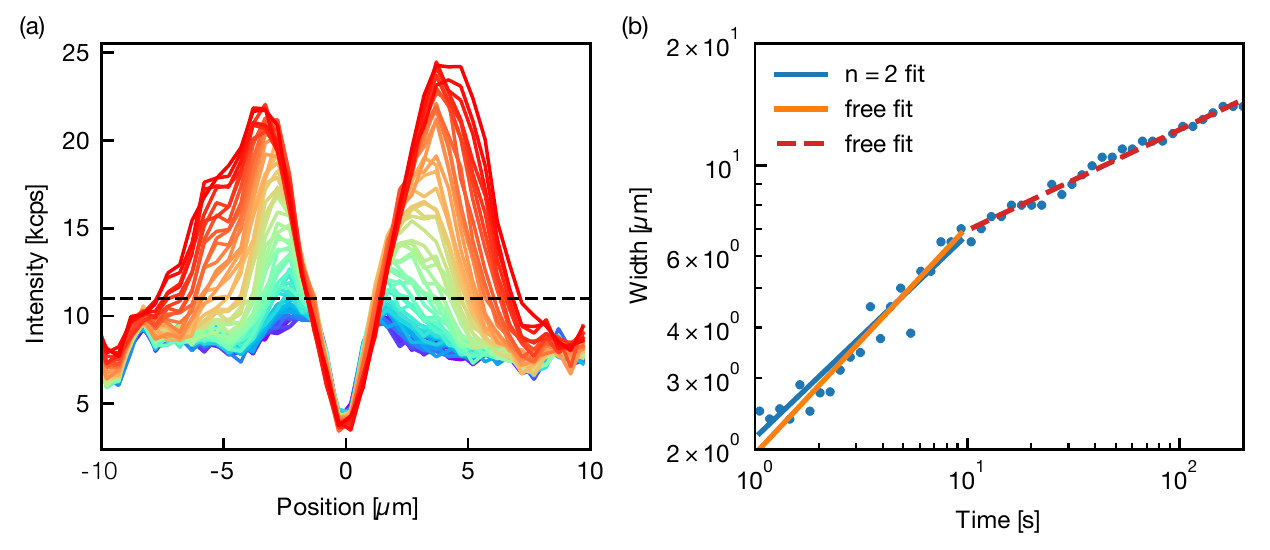}
  \caption{{\bf Time dependent measurement of \siv{} generation}. (a) Spatial distribution of \siv{} as a function of 532~nm (0.24~mW) illumination time. The duration was set from 0.3~s to 199.5~s with logarithmic spacing. The threshold for edge detection was set to 12 kcps. (b) Width of the \siv{} torus as a function of total 532~nm illumination time. Two different timescales are observed at early and late times. The data at different times was fitted with a model $\sigma(t) = D^{1/2}t^{1/2}$, where $t$ denotes the total 532~nm time, $D$ is a free coefficient, and n is the exponent describing the speed for the growth of the torus. The blue line is a fit to the data with $n$ constrained to 2. The yellow line is a fit to the early time data with $n$ as a free parameter, with $n = 1.8 \pm 0.1$. The red dashed line is a fit to the late time data with $n$ as a free parameter, with $n = 4.0 \pm 0.1$.  }
  \label{fig:Diffusion_Time_2}
\end{figure}

In addition to the strong absorption of holes from the \sivm{} centers, other impurities in the sample can also affect the carrier diffusion process. For example, in order to satisfy charge neutrality in the sample, positively charged defects (presumably positively charged substitutional nitrogen) should be present to compensate for the negative charges from \sivm{} prior to photoactivation of carriers. After generation of \siv{} centers via hole capture, one needs to consider the resulting space-charge potential from the remaining positively charged substitutional nitrogen and the itinerant electrons. This space-charge potential can affect the carrier diffusion significantly~\cite{Lozovoi2020}.

\subsection{Ionization dynamics of \siv{} centers}

\begin{figure}[h!]
  \centering
  \includegraphics[width = 129mm]{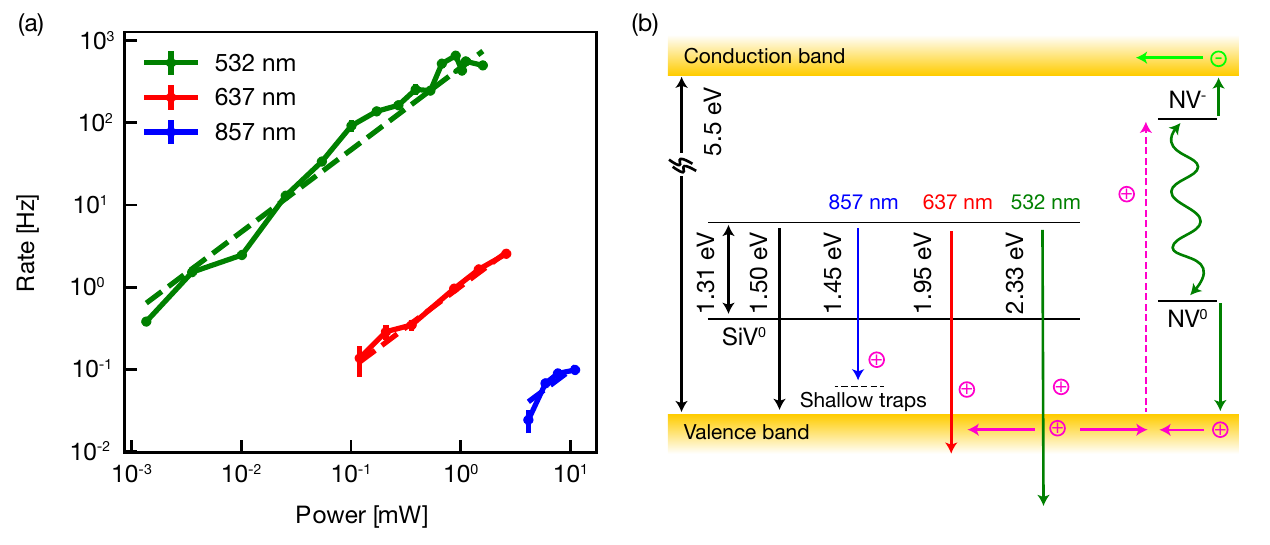}
  \caption{{\bf Ionization dynamics of \siv{} centers}. (a) Power dependent ionization rate of \siv{} centers at different wavelengths. The dashed lines are linear fits to the data. The fit for 637 nm is consistent with a linear scaling. The rates for 532~nm and 857~nm deviate from a linear scaling and saturate at higher powers. (b) Energy level diagram with the hypothesized microscopic model for \siv{} ionization at different wavelengths.}
  \label{fig:ionization}
\end{figure}

We probe the stability of \siv{} centers under different excitation wavelengths. The power dependence of the ionization rates are summarized in Fig.~\ref{fig:ionization}(a). We observe that the ionization rates can vary by orders of magnitude depending on the wavelength. When exciting at 857~nm, below \siv{} ionization threshold ($\sim$1.5~eV~\cite{Collins1995}), ionization is slow and the rate saturates at high powers. When exciting above the ionization threshold at 637~nm, a linear scaling of the ionization rate is observed, consistent with a single-photon ionization process. For 532~nm, the ionization rate is much faster compared to that of 637~nm, and the rate saturates at higher powers. We note that previous photoconducivity measurements on \siv{} showed relatively flat responses above 1.8~eV~\cite{Collins1995}. The difference in ionization rate between 532~nm and 637~nm excitation suggests influences of charge dynamics from other other coexisting defects.

The hypothesized ionization processes under different wavelengths are shown in Fig.~\ref{fig:ionization}(b). 

For 857 nm excitation, the excitation energy (1.45~eV) is below the ionization threshold of \siv{}, leading to a suppressed ionization rate. The non-negligible below-threshold ionization may be facilitated by shallow charge traps in the sample. These charge traps are photo-inactive, and have a finite lifetime for the trapped charges, which is independent to optical excitation. The holes from the \siv{} centers can tunnel to the charge traps, and the saturation of these traps will lead to the saturation of ionization. 

For 637~nm excitation, the excitation energy is higher than the ionization threshold, resulting in a single-photon ionization process with a linear scaling for the power dependence. 

With 532~nm excitation, photoinduced processes from the coexisting NV centers and P1 centers need to be considered, as the interplay between these processes and \siv{} ionization can modify the charge dynamics significantly. NV centers cycle between the negative charge state (NV$^-$) and the neutral charge state (NV$^0$) under 532~nm excitation, while this charge cycling is not possible under 637~nm excitation. NV$^-$ centers can capture the holes generated from photoionization of \siv{} efficiently due to their large hole-capture cross section \cite{Lozovoi2021}. Therefore, under 532 nm excitation NV centers can serve as a continuously replenished sink for holes, which leads to a faster instantaneous ionization rate. 

Additionally, the P1 center ionization threshold is around 1.7 eV, and the ionization rate is around ten times larger at 2.3 eV (532 nm) than at 1.9 eV (637 nm) \cite{Farrer1969,Nesladek1998,Isberg2006,Heremans2009}. The resulting positively charged P1 centers can then form a local space-charge potential \cite{Lozovoi2020}, preventing the diffusion of holes, resulting in a saturation of the ionization rate at high powers. 

We note that without detailed knowledge of the concentrations of different defects and the relevant capture and ionization rates for the photoinduced processes, it is difficult to disentangle the competing charge dynamics. A full model involving charge generation and transport, and local space-charge potential within the excitation volume may help to provide more a definitive understanding and assignment of the underlying processes, and would be an interesting avenue for future exploration.

\subsection{Comparing the photoactivation effect of 532 nm and 561 nm}
\begin{figure}[h!]
  \centering
  \includegraphics[width = 86mm]{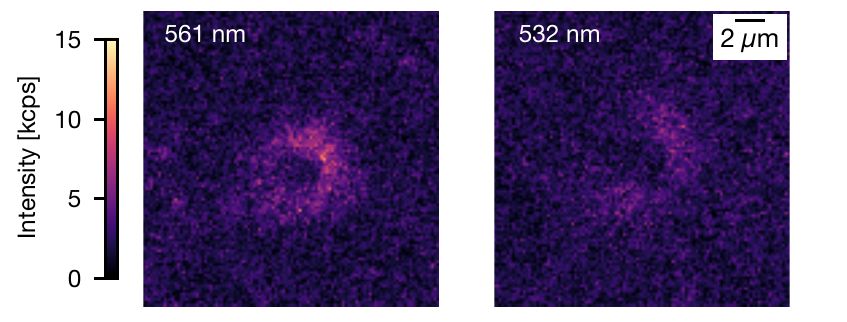}
  \caption{{\bf Stabilization of \siv{} with photoactivation of carriers using 561 nm illumination}. Optical illumination with 561 nm (left) and 532 nm (right) are focused on two locations separated by 30 microns. The illumination power is 0.46~mW and the illumination time is 30~s. The asymmetric pattern generated under 532 nm may arise from astigmatism of visible beam going through NIR optics at some tilt angle.}
  \label{fig:LA_561}
\end{figure}

To further confirm the origin of carriers in our sample, we performed fixed location illumination using 561 nm and 532 nm with the same illumination power and duration on two spots separated by 30 microns. Bright tori of \siv{} fluorescence of similar sizes are observed, as shown in Fig.~\ref{fig:LA_561}. Together with the wavelength dependence shown in the main text (Fig.~5), it can be concluded that a sharp transition for the photoactivation of holes exists between 561~nm and 595~nm. Notably, these wavelengths are to the blue and red, respectively, of the 575~nm zero-phonon line of NV$^0$, below which continuous charge state cycling of NV centers is possible. This sharp transition strongly suggests NV centers as the main source of itinerant carriers in our sample.

\subsection{Photogeneration of \siv{} using 637 nm excitation}
\begin{figure}[h!]
  \centering
  \includegraphics[width = 86mm]{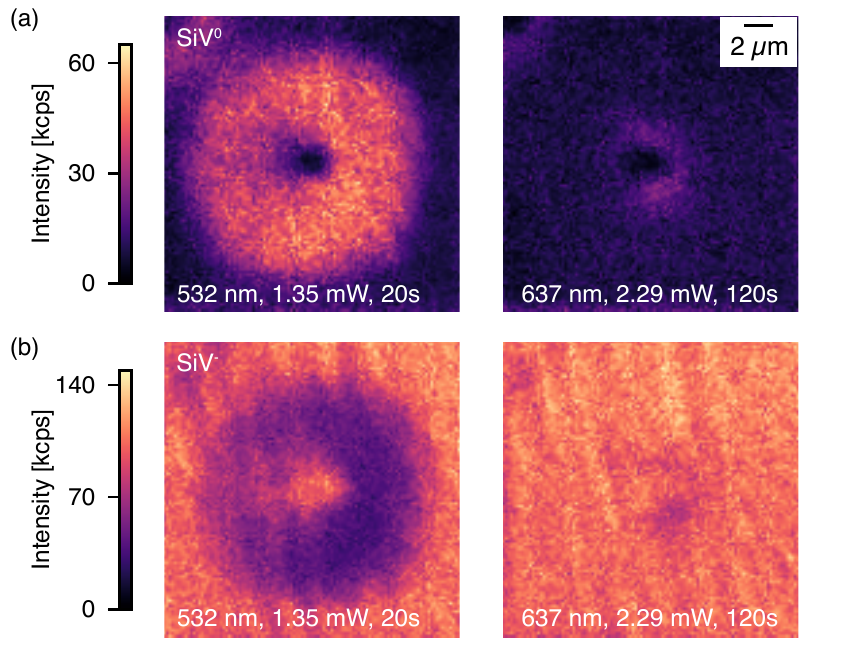}
  \caption{{\bf Stabilization of \siv{} with with 637 nm excitation}. (a) \siv{} spatial distribution after 532 nm laser illumination (left) or 637 nm laser illumination (right). (b) \sivm{} spatial distribution after 532 nm laser illumination (left) or 637 nm laser illumination (right). The whole area is initialized to be \sivm{} rich with low power 532 nm raster scans prior to the laser illumination. Despite the higher excitation power and the longer illumination time for 637 nm illumination, the photoactivation of carriers (inferred by photogeneration of \siv{}) is much weaker compared to that of 532 nm illumination.}
  \label{fig:LA_red_doping}
\end{figure}

In recent studies, it was observed that 638 nm excitation of \sivm{} centers can affect the charge state of nearby \sivm{} centers \cite{Gardill2021}. In sample D1, we observe a similar effect using 637~nm illumination, but the effect was less efficient comparing to that of 532~nm illumination (Fig.~\ref{fig:LA_red_doping}). After fixed location 637~nm illumination, we observe generation of a small amount of nonequilibrium \siv{} centers. At the same time, the spatial distributions of \siv{} and \sivm{} are inverted, suggesting that hole capture of \sivm{} is responsible for the stabilization of \siv{}. Under 637 nm illumination, the photoactivation of carriers cannot be accounted by the charge dynamics of NV centers, where NV$^-$ centers photoionize to the NV$^0$ centers and only produce a limited number of electrons. Similarly, substitutional nitrogen (N$_s^0$) photoionize weakly under 637~nm excitation, but during this process only electrons are generated \cite{Dhomkar2018}. With the high concentration of \sivm{} centers in our sample, it is likely that the photogeneration of nonequilibrium \siv{} under 637 nm illumination is related to continuous charge cycling of SiV centers \cite{Gardill2021}.

\subsection{Photogeneration of \siv{} in sample D2}
\begin{figure}[h!]
  \centering
  \includegraphics[width = 86mm]{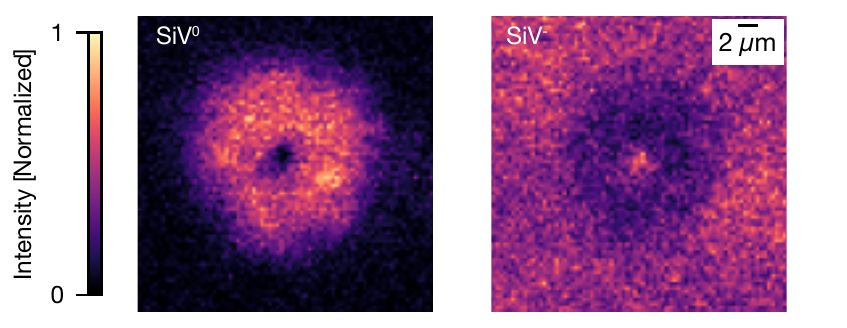}
  \caption{{\bf Charge state conversion of SiV centers with
  532~nm illumination on sample D2}. Spatial distribution of \siv{} centers (left) and \sivm{} centers (right) after a fixed location 532~nm illumination (0.98 mW for 60~s).}
  \label{fig:MT_confocal}
\end{figure}

To check that the photogeneration of \siv{} is not a unique phenomenon in a single sample, we repeat the photoactivation of itinerant carriers on another sample, D2. Similar to sample D1, we observe photogeneration of \siv{} and depletion of \sivm{} centers with fixed location 532~nm illumination~(Fig.~\ref{fig:MT_confocal}).

\begin{figure}[h!]
  \centering
  \includegraphics[width = 129mm]{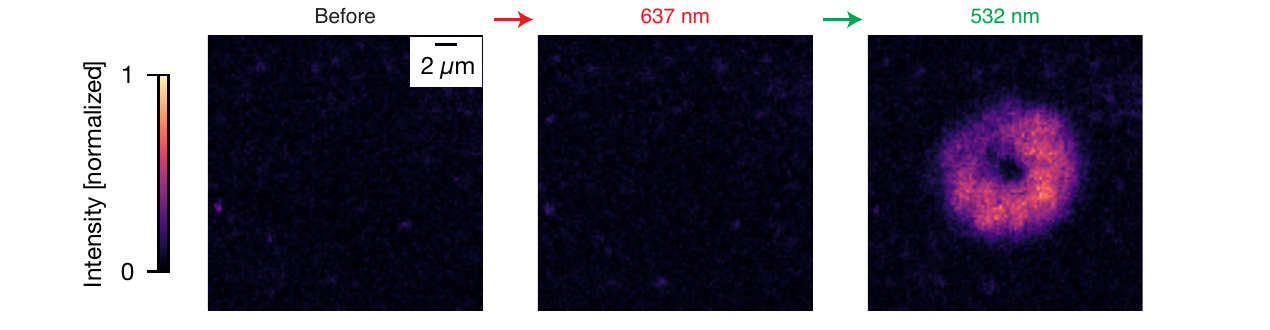}
  \caption{{\bf Wavelength dependence of photogeneration of \siv{} on sample D2}. Optical illumination at 637~nm and 532~nm are focused on the same location sequentially for 60~s with 0.68~mW. The arrows indicate the temporal sequence. \siv{} centers were only observed after 532 nm illumination.}
  \label{fig:MT_wavelength}
\end{figure}

In sample D2, under the same illumination condition as 532~nm, photogeneration of \siv{} centers cannot be observed with 637~nm illumination, consistent with the observation in sample D1 (Fig.~\ref{fig:MT_wavelength}). Considering the similar NV concentration on the two samples, this also suggests NV center as the source of holes on sample D2.

\begin{figure}[h!]
  \centering
  \includegraphics[width = 86mm]{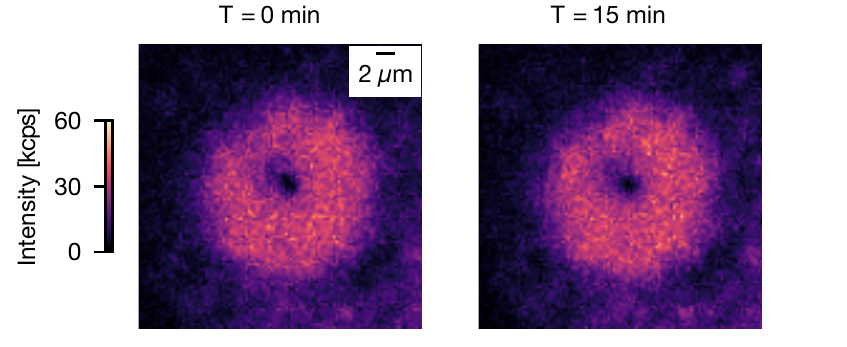}
  \caption{{\bf Charge state stability in the dark on sample D2}. After fixed location 532~nm illumination, nonequilibrium \siv{} centers are probed after a dark period of 15 minutes with 0.29~mW of 857 nm. The spatial distribution of \siv{} remains stable.}
  \label{fig:MT_dark}
\end{figure}

We probe the stability of the photogenerated \siv{} centers in the dark in sample D2. No appreciable change of population is observed after 15 minutes (Fig.~\ref{fig:MT_dark}), suggesting the nonequilibrium \siv{} centers in this sample are also long-lived.

\subsection{Photogeneration of \siv{} in sample D3}
In this section, we show that the charge state dynamics of SiV can be strongly sample dependent by studying the photo-dynamics on a third sample, D3.

\begin{figure}[h!]
  \centering
  \includegraphics[width = 86mm]{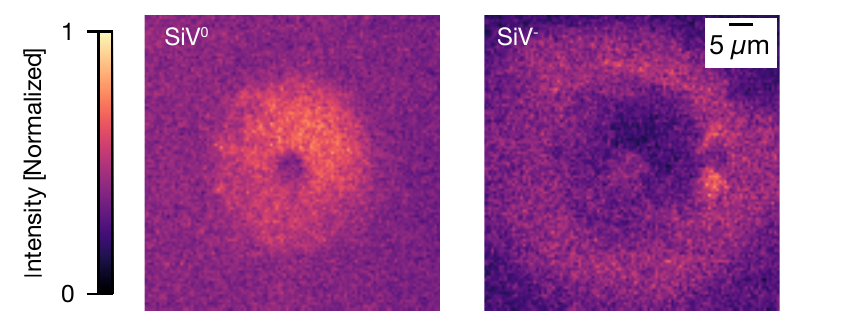}
  \caption{{\bf Charge state dynamics of SiV centers in sample D3}. Spatial distribution of \siv{} (left) and \sivm{} (right) after fixed location illumination using 1 mW 532 nm for 60s. The \siv{} signal shows a bright torus around the illumination position. The \sivm{} signal shows an inner dark torus with the same size as the \siv{} torus as well as a larger bright torus. The readout of \siv{} was using 8~mW 857 nm excitation while the readout of \sivm{} was using 0.37~mW 637~nm.}
  \label{fig:SiV2m}
\end{figure}

532 nm illumination is focused at a fixed location on sample D3. Afterwards, the spatial distribution of \siv{} and \sivm{} centers are probed~(Fig.~\ref{fig:SiV2m}). For \siv{}, we observe appearance of a bright torus (Fig.~\ref{fig:SiV2m}, left), similar to the observation in sample D1 and D2. However, the spatial distribution of \sivm{} is drastically different (Fig.~\ref{fig:SiV2m}, right). Four salient features are worth noting: (1) \sivm{} is bright under direct 532~nm illumination; (2) near the illumination location, a dark torus can be observed, the size of which is consistent with the bright torus of \siv{}; (3) an additional bright torus is observed outside the smaller dark torus; (4) far away from the illumination location, the SiV centers are neither in the neutral charge state or in the negative charge state.

The first two features are consistent with our observations in sample D1 and D3, while the last two features are drastically different. First, the spatial distributions of \siv{} and \sivm{} are no longer inverted. At the same time, both \siv{} and \sivm{} are dark far away from the illumination location. Since optical illumination cannot create SiV centers but can only modify the SiV charge state, this suggests that a third charge state of SiV is present in this sample. Second, outside the dark torus of \sivm{}, a bright torus corresponding to photogeneration of \sivm{} is observed. Without any initial \siv{} population, the photogeneration of \sivm{} centers suggests that the initial state prior to \sivm{} photogeneration is SiV$^{2-}$. In sample D3, the SiV centers are thermodynamically stable in the form of SiV$^{2-}$. With the photoactivated holes, the SiV$^{2-}$ centers are first converted to \sivm{} centers via single hole capture. Afterwards, the photogenerated \sivm{} centers can be converted to \siv{} centers via additional hole capture. The larger torus size for \sivm{} compared to \siv{} can then be explained by the fact that photogenenration of \siv{} is possible only after photogeneration of \sivm{}.

Based on our observations, the following model can account for the differences among these samples: the hole capture effect of SiV centers is robust across different samples, while the preferred charge state in equilibrium can vary depending on the details of the samples (\sivm{} in sample D1 and D2, SiV$^{2-}$ in sample D3). This preferential charge state of SiV centers depends on the local Fermi level of the diamond \cite{Gali2013}, where a higher Fermi level favors SiV$^{2-}$ and lowering the Fermi level favors \sivm{} and eventually \siv{}. With a lower concentration of nitrogen, the Fermi level is closer to the middle of the bandgap, favoring more \sivm{}. With a higher concentration of nitrogen, the Fermi level is pinned to 1.7~eV below the conduction band minimum, favoring the stabilization of SiV$^{2-}$. For samples D1, D2 and D3, we are unable to measure the difference of nitrogen concentrations using FTIR due to the limited sensitivity. Additionally, charge traps in the samples can affect the effect of Fermi level pinning from nitrogen. Nevertheless, the observation of different thermodynamically preferred charge state in different samples may resolve the dispute in previous literatures about the dark state of \sivm{} centers \cite{Gardill2021,Dhomkar2018} .

We note that the above discussion seems to resolve the apparent discrepancies in the charge capture process among samples, but the SiV charge state under direct 532~nm illumination remains unresolved. Specifically, in our samples, \sivm{} centers are bright under 532~nm excitation, while in some other works, \sivm{} centers are reported to enter a dark state under 515~nm or 532~nm excitation \cite{Gardill2021,Dhomkar2018}. More work is needed to resolve this discrepancy, but again, the concentration of local defects can play a big role. For example, even within a diffraction-limited spot, optical excitation can still address many SiV centers, NV centers and P1 centers simultaneously, carrier transport between these defects needs to be considered. The different concentration for these three defects in different samples may be used to resolve the discrepancy of \sivm{} dynamics under direct 532~nm illumination.

\subsection{Single center charge dynamics with 532~nm illumination}
\begin{figure}[h!]
  \centering
  \includegraphics[width = 86mm]{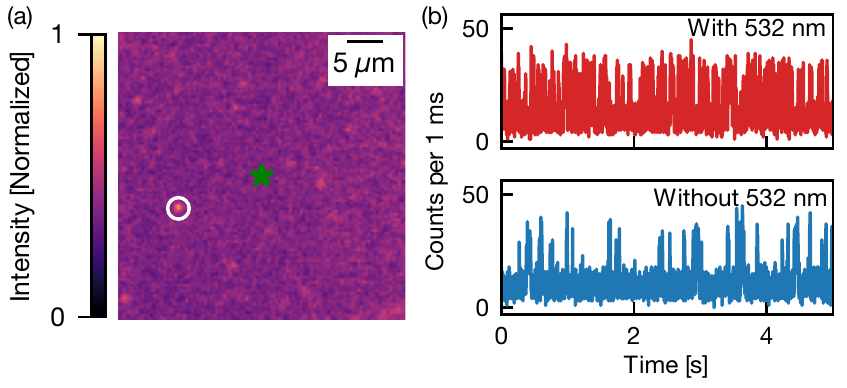}
  \caption{{\bf Effect of remote 532~nm illumination on single centers}. (a) Fluorescence scan on sample D4 showing individual bright centers. The green star indicates the position for continuous 532 nm illumination. The bright spot in the white circle indicates the single center studied in (b). (b) Fluorescence count rate of the single center under 857~nm (7~mW) with and without the 532~nm illumination (0.2~mW). The distance between the single center and the illumination location is $\sim$12.5 $\mu$m.}
  \label{fig:AK_single}
\end{figure}

In this section, we present preliminary measurements of the influence of photoactivated carriers on the charge state dynamics of individual SiV centers. The single centers measured in this section are from two samples: (1) a region with low silicon doping in sample D1 and (2) an additional silicon implanted sample D4. Due to the low density of centers, single centers are resolvable as individual bright spots in the fluorescence scan~(Fig.~\ref{fig:AK_single}(a)). 

\begin{figure}[h!]
  \centering
  \includegraphics[width = 129mm]{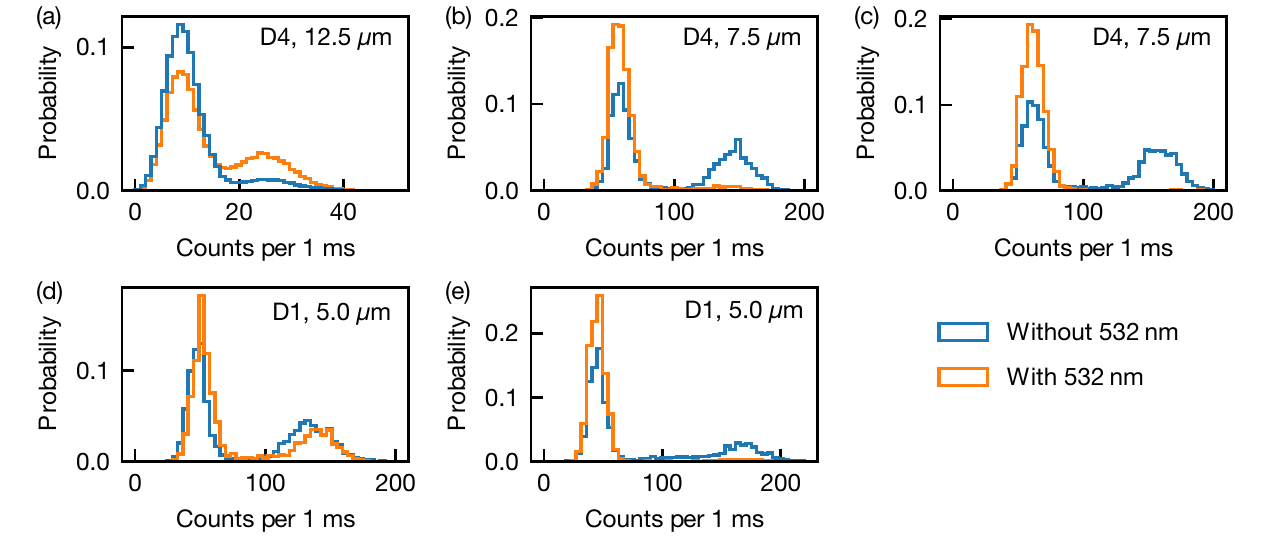}
  \caption{{\bf Variation of charge state dynamics with remote 532~nm illumination for single centers}. Histograms of the fluorescence time traces on 5 different single centers on sample D1 and D4 with and without remote 532~nm illumination are measured. The label in each plot denotes the distance between the 532~nm illumination and the single center. The 532~nm power was kept 1.5~mW except for (a) where 0.2~mW was used. The lower count rate for the center in (a) was due to the usage of a different detector with lower efficiency. (a) shows the histogram for the single center studied in Fig.~\ref{fig:AK_single}.}
  \label{fig:single_hist}
\end{figure}

For the single centers in these two samples, the emission of centers show visible telegraph switching between a bright state and a dark state (Fig.~\ref{fig:AK_single}(b)). The high contrast charge state switching allows for quantification of the charge state stability. For these single centers, we take fluorescence time traces with and without continuous 532~nm illumination in a nearby location. The emission statistics of several centers change upon remote photoactivation (Fig.~\ref{fig:single_hist}). For some centers, we observed a significant decrease of population in the bright state, while the opposite change was observed for a center in sample D4 (Fig.~\ref{fig:single_hist}(a)). Additionally, no appreciable change was observed for one of the centers in sample D1 (Fig.~\ref{fig:single_hist}(d)). The variation among single centers may arise from an unknown source of local inhomogeneity; for example, high strain \cite{Rose2018} may modify the charge capture process.

\clearpage
\bibliography{doping}